\documentclass[%
 aip,
 amsmath,amssymb,
 reprint,%
]{revtex4-1}

\usepackage{graphicx}
\usepackage{dcolumn}
\usepackage{bm}

\usepackage[utf8]{inputenc}
\usepackage[T1]{fontenc}
\usepackage{mathptmx}
\usepackage{etoolbox}

\makeatletter
\def\@email#1#2{%
 \endgroup
 \patchcmd{\titleblock@produce}
  {\frontmatter@RRAPformat}
  {\frontmatter@RRAPformat{\produce@RRAP{*#1\href{mailto:#2}{#2}}}\frontmatter@RRAPformat}
  {}{}
}%
\makeatother
\begin{document}

\preprint{AIP/123-QED}

\title{Evaporative cooling and deposition patterns of evaporating $Al_2O_3$ nanofluid droplets}
\author{S.K. Saroj*}
\email{sunilkrsiitk@gmail.com}.
 \altaffiliation{Department of mechanical engineering, Indian Institurte of technology Kanpur, Kanpur, U.P, India 208016.}
\author{P.K. Panigrahi}%
\altaffiliation{Department of mechanical engineering, Indian Institurte of technology Kanpur, Kanpur, U.P, India 208016.}

\date{\today}

\begin{abstract}
The present study investigates evaporative cooling and the resulting deposition patterns of a sessile droplet of $Al_2O_3$-based nanofluid evaporating on a hydrophobic glass substrate maintained at different temperatures. A goniometric arrangement is employed to record the side-view evolution of droplet geometry during evaporation. Infrared (IR) thermography is used to measure the interfacial temperature distribution, while confocal microscopy is utilized to visualize top view image and resulting deposition pattern. In addition, $\mu$-PIV measurements are performed to quantify the internal velocity field. The evaporation predominantly occurs in the pinned contact line mode for both heated and non-heated substrates, with only slight recession observed in the non-heated case. The droplet height and contact angle decrease linearly with time, and scaling relations are proposed to describe the evolution of droplet geometry and volume. A non-dimensional parameter $\Pi_{rel}$ is introduced to quantify the transition in the deposition pattern. For $\Pi_{rel} \le$ 1, a framework of interconnected irregular polygonal tessellation network structures is observed at the peripheral region for $T_s \le 26^\circ$C. Possibly, this type of unique structure at the peripheral region is observed for the first time in evaporating droplets. This polygonal network structure is suppressed with substrate heating, leading to the formation of a classical coffee-ring pattern for  $ 1 <\Pi_{rel} \le$ 10. Above a critical substrate temperature ($T_s > 40^\circ$C), dual-ring formation along with central particle deposition is observed for $\Pi_{rel} >$ 10. The interfacial temperature distribution shows higher temperatures near the contact line and lower temperatures toward the droplet apex, and a universal scaling for the interfacial temperature profile is proposed. The derived temperature distribution inside the droplet exhibits a linear profile, indicating heat transfer dominated by conduction. The internal velocity increases with increasing substrate temperature and exhibits asymmetric flow structures with multiple vortices. Evaporative cooling intensifies with increasing surface temperature. The evaporation flux increases with substrate temperature, leading to an increase in capillary velocity. Appropriate scaling of evaporation flux and capillary flow is established. The observed dynamics are attributed to thermo-capillary flow induced by evaporative cooling, which enhances internal circulation, evaporation flux, and ultimately modifies nanoparticle deposition morphology. The underlying physical mechanism can be summarized as
\begin{equation}
T_s \uparrow \Rightarrow 
J \uparrow \Rightarrow 
(T_s - T_{int}) \uparrow \;(\text{evaporative cooling}) \Rightarrow 
(T_{edge}-T_{apex}) \uparrow \Rightarrow 
Ma \uparrow \Rightarrow 
U_c \uparrow \Rightarrow 
\text{modified deposition pattern}.
\end{equation}

where $T_s$, $T_{int}$, $T_{edge}$, $T_{apex}$, $J$, $Ma$, and $U_c$ denote the substrate temperature, interfacial temperature, temperature at the contact line (edge), temperature at the droplet apex, evaporation flux, Marangoni number, and capillary velocity, respectively.

\end{abstract}

\maketitle

\section{Introduction}
Droplets are ubiquitous in nature and frequently encountered in everyday life. For example, early in the morning small droplets can often be observed on plant leaves and grasses due to condensation. The geometrical shape and spreading behavior of a droplet on a surface are governed by the wettability characteristics of the substrate. As the sun rises and the ambient temperature increases, these droplets gradually disappear due to evaporation. Such simple natural observations have motivated extensive scientific investigations into the evaporation dynamics of droplets. Understanding droplet evaporation is crucial for a wide range of technological applications, including spray cooling, spray coating, inkjet printing, agricultural spraying, and pattern formation during drying. Droplet evaporation is particularly important in thermal management applications such as spray cooling\cite{mollaret2004experimental}, where a large number of droplets are deposited on heated surfaces.

The evaporation of a sessile droplet on a substrate occurs in three distinct modes: constant contact radius (CCR), constant contact angle (CCA), and a mixed mode involving both partially pinned and partially moving contact lines \cite{gelderblom2011water,Saroj2019Drying,Saroj2020Magnetic,Dash2014,Nguyen2018,nguyen2012lifetime}. During evaporation, droplets absorb latent heat from the substrate, thereby reducing its temperature. The efficiency of this cooling process strongly depends on the evaporation dynamics, which are influenced by parameters such as substrate temperature, droplet geometry, and heat and mass transfer mechanisms \cite{Nguyen2018,gatapova2014evaporation,Hu2002}. Nanofluids are suspensions of nanosized particles dispersed in a base liquid and exhibit enhanced thermal properties such as thermal conductivity, viscosity, and heat capacity compared to the base fluid\cite{Vajjha2008,Sridhara2011,Nindo2007,Esfe2014}. This optimized thermo-physical properties are significantly affects the evaporation dynamics and most importantly accelerate the evaporation process via faster conducting the heat flow from the surface to interface of the droplet. When a droplet contains suspended or dissolved solutes, the evaporation process becomes more complex because the internal flow within the droplet redistributes the solute particles\cite{saroj2018effect}. This transport leads to self-assembly and the formation of various deposition patterns after complete evaporation. The final deposit morphology depends on several factors such as evaporation rate, internal capillary flow, Marangoni flow, particle interactions, and substrate properties \cite{Bhardwaj2009,Dash2014,Hu2002}. A colloidal droplet on a solid substrate often produces a non-uniform deposition pattern after complete drying\cite{Xu2014}. This non-uniform deposition pattern is typically ring-shaped and is commonly referred to as the coffee-ring effect. The formation of the coffee ring is mainly attributed to the pinning of the contact line and the radial outward capillary flow during evaporation \cite{Deegan1997,Deegan2000,zhao2018nanoparticle,Hu2002}. These patterns are important for many industrial and biological processes, including coating technologies, surface micropatterning, DNA molecular stretching, nanophotonics, nanoelectronics, magnetoelectronics, biochemical sensing, and medical diagnostics \cite{Parsa2015,Zhong2014}. The evaporation dynamics and resulting drying patterns depend on the size, shape, and type of nanoparticles present in the fluid. Shin et al.~\cite{Shin2014} reported the evaporation dynamics and deposition pattern formation of Al$_2$O$_3$ (50 nm diameter) nanofluid droplets on a hydrophilic glass substrate as a function of nanoparticle concentration. They observed a reduction in the total evaporation time and the initial contact angle, which was attributed to the decrease in surface tension with increasing nanoparticles concentration.

Marangoni flow is another important mechanism that can significantly influence the drying pattern \cite{Hu2006}. The surface tension gradient along the liquid–air interface arises due to non-uniform temperature distribution caused by evaporative cooling. This gradient induces Marangoni convection, which drives the fluid from regions of lower surface tension to regions of higher surface tension \cite{Hu2006,Hu2005,Ristenpart2007}. Askounis et al.~\cite{Askounis2017} experimentally investigated the effect of localized heating (at the edge and the center beneath the droplet) on droplet evaporation and Marangoni flow using infrared thermography and optical imaging. They observed that edge heating produces vortices on the opposite side of the heating location, whereas center heating generates twin vortices inside the droplet. Local heating was also found to promote stick–slip motion of the contact line during evaporation.


Parsa et al.~\cite{Parsa2015} investigated the effect of substrate temperature on the deposition pattern formation of water–CuO nanofluid droplets on smooth silicon wafers. This study mainly focused on the post-evaporation process. Similarly, Patil et al.~\cite{Patil2016} studied the influence of substrate temperature, substrate wettability, and particle concentration on the deposition patterns formed by evaporating water droplets containing polystyrene latex beads with a diameter of 0.46~$\mu$m. Furthermore, the coupling among heat transfer, internal flow, particle transport, and diffusion mechanisms plays an important role in temperature-dependent evaporation, the resulting cooling behavior, and the deposition structures of nanoparticles on the substrate. Therefore, it is crucial to investigate the evaporation dynamics, internal flow behavior, interfacial temperature variation, and the mechanisms governing evaporative cooling and the resulting deposition morphology in detail.

The present study reports a systematic investigation of the evaporation dynamics of droplets containing Al$_2$O$_3$ nanoparticles on a hydrophobic substrate at temperatures below, equal to, and above the ambient temperature. 
The measurement of the geometrical evolution, interfacial temperature distribution, and internal flow dynamics of the droplet, while also visualizing the top-view temporal evolution of the droplet surface and the resulting final deposition morphology. The temporal evolution of droplet height and contact angle is scaled with evaporation time, leading to a relation describing the geometrical evolution of the droplet during evaporation. Furthermore, appropriate scaling arguments are developed to explain the observed evaporation behavior and to establish correlations between the governing parameters. The measured interfacial temperature is highest at the contact line and decreases toward the apex region. A good collapse of the interfacial temperature for different substrate heating conditions is observed; therefore, a universal interfacial temperature profile is proposed as a function of radial position. The relevant non-dimensional numbers, such as the Marangoni number, and Rayleigh number, capillary number, and jakob number are evaluated to identify the dominant transport mechanisms during evaporation. A critical substrate temperature ($T_s$ = 40 $^\circ$C) is identified at which a transition in the deposition behavior occurs. For non-heated and cooled substrates, a framework of interconnected irregular polygonal tessellation network structures at the peripheral region is observed. These structures are suppressed at higher substrate temperatures, leading to the formation of dual-ring structures followed by central particle deposition. Radial cracks are also observed at higher substrate temperatures due to the reduction in deposit width. In addition, the evaporative cooling performance is quantified and the heat-transfer effectiveness is evaluated to determine the cooling efficiency at different substrate temperatures. The cooling effectiveness is calculated and shows it is almost constant, $\epsilon$ $\sim$ 1.68 for all heating substrate temperature. The $\epsilon$ = 1.68 indicates higher evaporative cooling compared to the heat conduction through the droplet. Additionally, the non-dimensional Jakob number, $J_a <$ 1 shows the evaporation dominance over the heat conduction through the droplet.

The present work provides new insights into the coupling between temperature-dependent evaporation, internal flow dynamics, and nano-particle deposition, which are important for applications in spray cooling and coating technologies.

\section{Experimental Details}

\subsection{Solution and Substrate}

The water-based nanofluid was prepared by dispersing Al$_2$O$_3$ nanopowders (Sigma-Aldrich, average diameter 13~nm) in deionized water at a mass concentration of 1.0\%. The solution was placed in an ultrasonic bath prior to the experiments to ensure uniform dispersion of the nanoparticles. The glass substrate used in the experiments was treated with octadecyltrichlorosilane (OTS) solution to render the surface hydrophobic. The glass shape is circular with diameter equal to 20 mm and thickness is equal to 100 $\mu$m. This glass surface is transformed in to the  hydrophobic glass surface in two steps. In the first step, the glass surface was cleaned with acetone and subsequently immersed in piranha solution for 15~min. The substrate was then removed, rinsed alternately with water and acetone several times, and dried in air. In the second step, the cleaned glass substrate was immersed in the OTS solution for 4~h at a temperature range of 20--25$^\circ$C. After treatment, the substrate was rinsed alternately with toluene and ethanol multiple times, followed by washing with an ethanol--water mixture and finally with water. The substrate was dried in air and then placed in an oven at 100$^\circ$C.
The initial contact angle of a water droplet on the treated substrate was measured to be $96 \pm 1^\circ$. The root-mean-square (RMS) surface roughness of the glass substrate was 3.5~nm. Fluorescent particles of diameter 2.1~$\mu$m (Invitrogen) were dispersed in the nanofluid as tracer particles for $\mu$-PIV measurements. A measured volume of $(2.0 \pm 0.05)\,\mu$L nanofluid droplet was deposited on the substrate.The thermophysical properties of the nanofluid used in the present study are provided in Table~\ref{tab:nanofluid_properties}. The ambient temperature and relative humidity were 26$^\circ$C and 50 \%, respectively.

\begin{figure*}[htb!]
    \centering
    \includegraphics[width=1\textwidth]{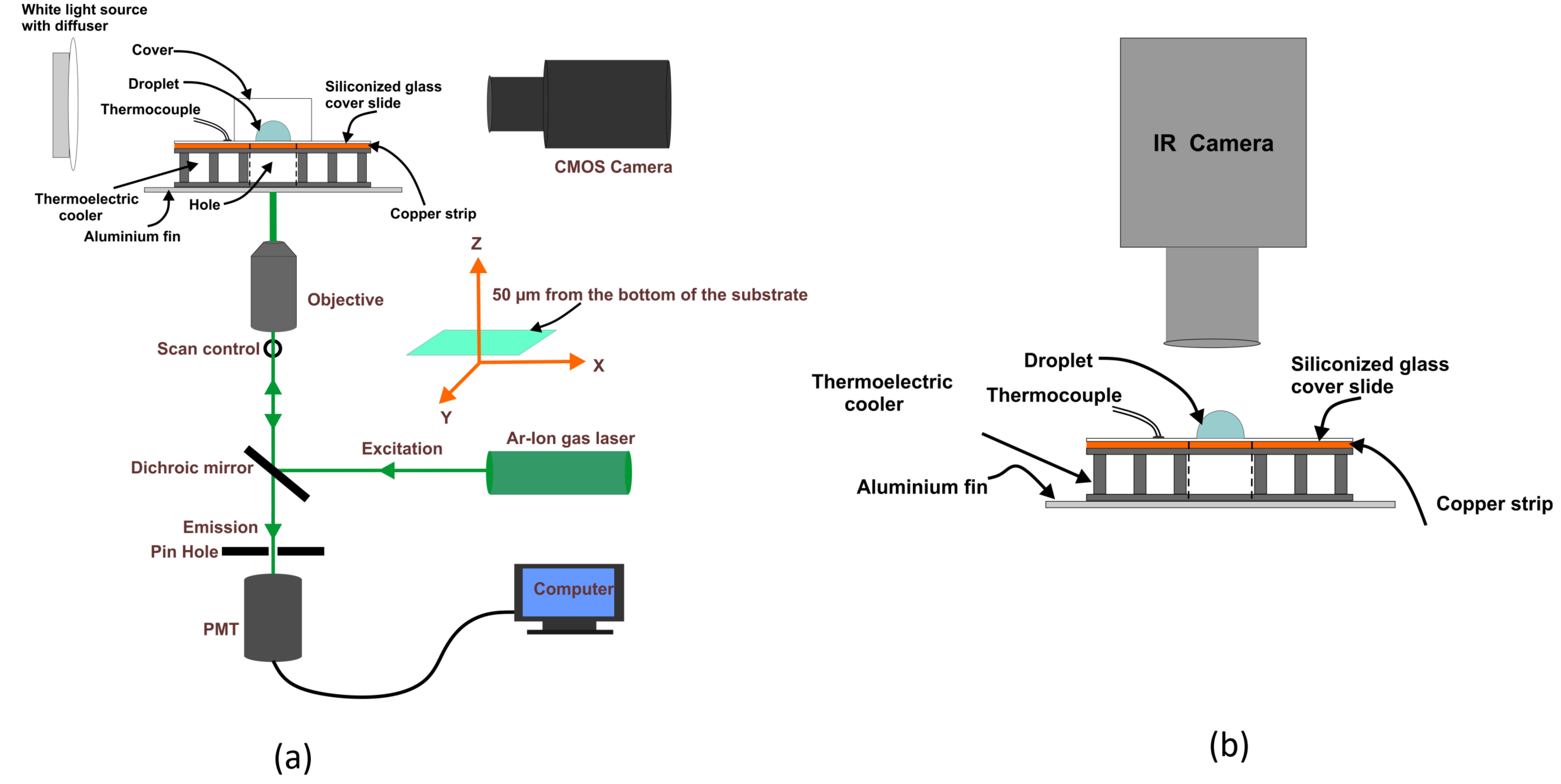}
    \caption{(a) Experimental setup for side view visualisation and particle motion inside the droplet, (b) Infrared thermography imaging arrangement.}
    \label{fig:setup}
\end{figure*}

\begin{figure}
    \centering
    \includegraphics[width=1\linewidth]{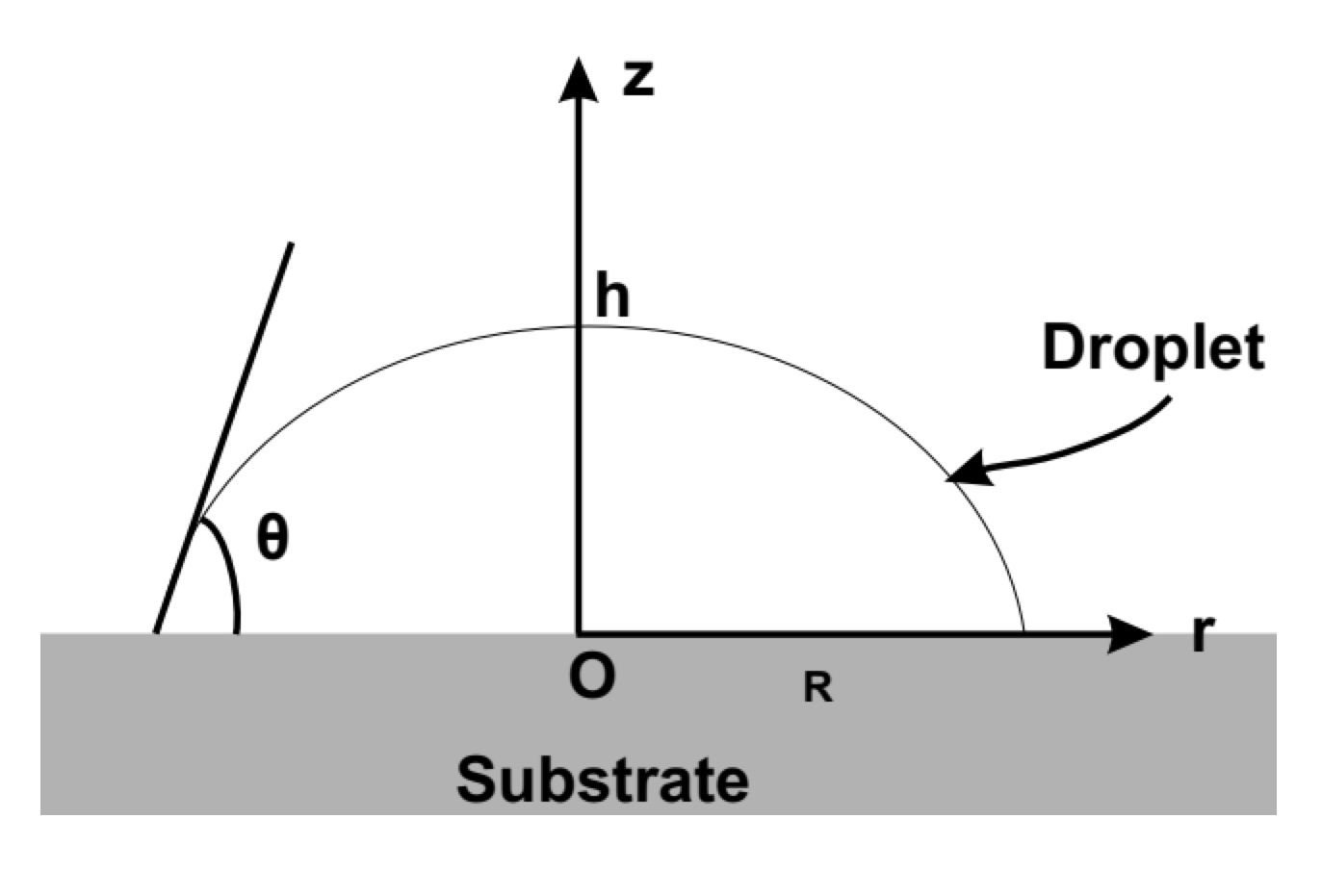}
    \caption{A sketch showing the geometrical parameter of the droplet}
    \label{fig:sketch}
\end{figure}


\renewcommand{\arraystretch}{2.5} 
\begin{table*}[htb!]
\centering
\caption{Thermophysical properties of water, Al$_2$O$_3$ nanoparticles, and Al$_2$O$_3$--water nanofluid at 25$^\circ$C. Here, $\phi$ denotes the nanoparticle volume fraction. Subscripts $w$, $p$, and $nf$ represent water, particle, and nanofluid, respectively, and $T$ is the temperature in Kelvin.}

\resizebox{\textwidth}{!}{
\begin{tabular}{|l|l|l|l|l|}
\hline
Property & Water & Particle (Al$_2$O$_3$) & Nanofluid relation & Nanofluid property \\
\hline

Density ($\rho$) 
& $\rho_w = 997~\mathrm{kg\,m^{-3}}$ 
& $\rho_p = 3970~\mathrm{kg\,m^{-3}}$ 
& $\rho_{nf}=(1-\phi)\rho_w+\phi\rho_p$ \cite{PakCho1998}
& $\rho_{nf}=1005~\mathrm{kg\,m^{-3}}$ \cite{CRC2014} \\
\hline

Viscosity ($\mu$) 
& $\mu_w = 8.9 \times10^{-4}$ $\mathrm{Pa\,s}$ \cite{Fox2011}
& --- 
& $\mu_{nf}=\mu_w(1+2.5\phi)$ \cite{Einstein1906}
& $\mu_{nf} = 8.97\times10^{-4}~\mathrm{Pa\,s}$ \cite{CRC2014} \\
\hline

Specific heat ($c_p$) 
& $c_{p,w} = 4182~\mathrm{J\,kg^{-1}K^{-1}}$ 
& $c_{p,p} = 765~\mathrm{J\,kg^{-1}K^{-1}}$ \cite{CRC2014} 
& $(\rho c_p)_{nf}=(1-\phi)(\rho c_p)_w+\phi(\rho c_p)_p$ \cite{XuanRoetzel2000}
& $c_{nf} = 4131~\mathrm{J\,kg^{-1}K^{-1}}$ \\
\hline

Thermal conductivity ($k$) 
& $k_w = 0.60~\mathrm{W\,m^{-1}K^{-1}}$ 
& $k_p = 36~\mathrm{W\,m^{-1}K^{-1}}$ 
& $k_{nf}=k_w\frac{k_p+2k_w-2\phi(k_w-k_p)}{k_p+2k_w+\phi(k_w-k_p)}$ \cite{Maxwell1873}
& $k_{nf} = 0.617~\mathrm{W\,m^{-1}K^{-1}}$ \cite{CRC2014} \\
\hline

Thermal expansion ($\beta$) 
& $\beta_w = 2.57\times10^{-4}~\mathrm{K^{-1}}$ 
& $\beta_p = 8\times10^{-6}~\mathrm{K^{-1}}$ 
& $\beta_{nf}=(1-\phi)\beta_w+\phi\beta_p$ \cite{PakCho1998}
& $\beta_{nf} = 2.0947\times10^{-4}~\mathrm{K^{-1}}$ \cite{CRC2014} \\
\hline

Thermal diffusivity ($\alpha$) 
& --- 
& --- 
& $\alpha_{nf}=\dfrac{k_{nf}}{(\rho c_p)_{nf}}$ \cite{Incropera2007}
& $\alpha_{nf} = 1.49\times10^{-7}~\mathrm{m^2\,s^{-1}}$ \cite{Incropera2007} \\
\hline
\end{tabular}
}

\label{tab:nanofluid_properties}
\end{table*}

\subsection{Experimental Arrangement}

Fig~\ref{fig:setup} (a) illustrates the experimental setup. A cylindrical hollow thermoelectric cooler (TE Technology Inc., Model TE-8-0.45-1.3) was used to heat the substrate to the desired temperature. A circular copper plate with a 3~mm diameter hole at the center was attached beneath the glass substrate. The copper plate was connected to the thermoelectric cooler using thermal grease to ensure efficient heat conduction. A thermocouple was placed on the glass substrate to monitor the temperature. The experiments performed when the thermocouple gives the constant reading of the desired temperature for around 5 minute. Thereafter, a droplet of known volume was deposited onto the surface using a micropipette. A goniometric arrangement was employed to record instantaneous side-view images during evaporation. The goniometric system consisted of a CMOS camera and a light source equipped with a diffuser. The diffuser minimized heat transfer from the light source, preventing interference with the evaporation process. Confocal microscopy was used to visualize particle motion, top-view droplet surface and deposition patterns. The time interval between two consecutive images was set to 1.29~s. A laser source with a wavelength of 488~nm was used to excite the fluorescent particles. A photomultiplier tube (PMT) captured the emitted fluorescence signal. The pinhole in the confocal system minimized background noise. The velocity vector field was determined using PIV processing with DynamicStudio software. Adaptive cross-correlation was employed for the PIV analysis. The interrogation window size and overlap were set to $32 \times 32$ pixels and 25\%, respectively. A infrared imaging system is employed to measure the liquid-air interface temperature during evaporation as shown in Fig.~\ref{fig:setup} (b). The experimental uncertainties of ambient temperature, relative humidity, substrate temperature, contact angle, and droplet contact diameter were $\pm 1^\circ$C, $\pm 3\%$, $\pm 0.5^\circ$C, $\pm 3^\circ$, and $\pm 0.09$~mm, respectively.

\section{Results}
 This section presents the experimental results and the derived equations based on scaling analysis.
\subsection{Evaporation Characteristics}

The present study investigates the evaporation dynamics of a sessile colloidal droplet containing $Al_2O_3$ nanoparticles placed on a heated hydrophobic glass substrate. Fig.~\ref{fig:sketch} illustrates the geometrical parameters of the droplet. Experiments are conducted at different substrate temperatures, $T_s$ = 22, 26, 40, 53 and 65$^\circ$C. The ambient temperature during the experiments was maintained at 26$^\circ$C. Therefore, the evaporation process was examined under both cooling ($T_s < T_{amb}$) and heating ($T_s > T_{amb}$) conditions of the substrate.

\begin{figure*}[htb!]
    \centering
    \includegraphics[width=1\textwidth]{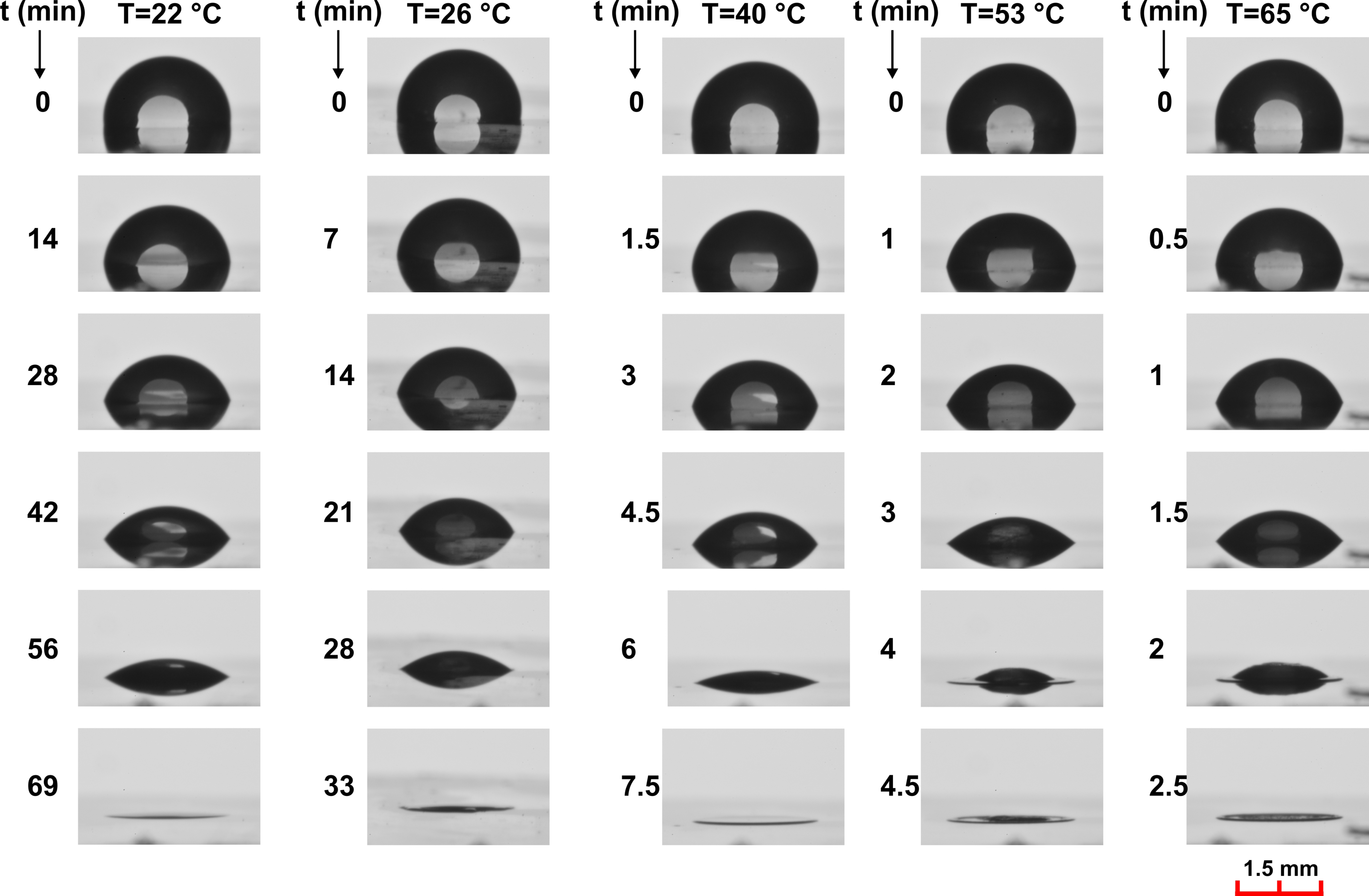}
    \caption{Side-view images of the evaporating sessile droplet at different substrate temperatures. The total evaporation time decreases with increasing $T_s$, and interfacial deformation becomes prominent at higher temperatures.}
    \label{fig:camera}
\end{figure*}

Fig.~\ref{fig:camera} shows the side-view images of the evaporating droplet at different substrate temperatures. It is observed that the total evaporation time decreases with increasing substrate temperature. At higher substrate temperatures ($T_s \geq 40^\circ$C), a noticeable deformation over the droplet interface appears, particularly during late stage of evaporation at lower contact angles. This interfacial deformation is likely caused by thermally induced Marangoni stresses along the liquid–air interface arising from temperature gradients along the droplet interface.

\begin{figure*}[htb!]
    \centering
    \includegraphics[width=1\textwidth]{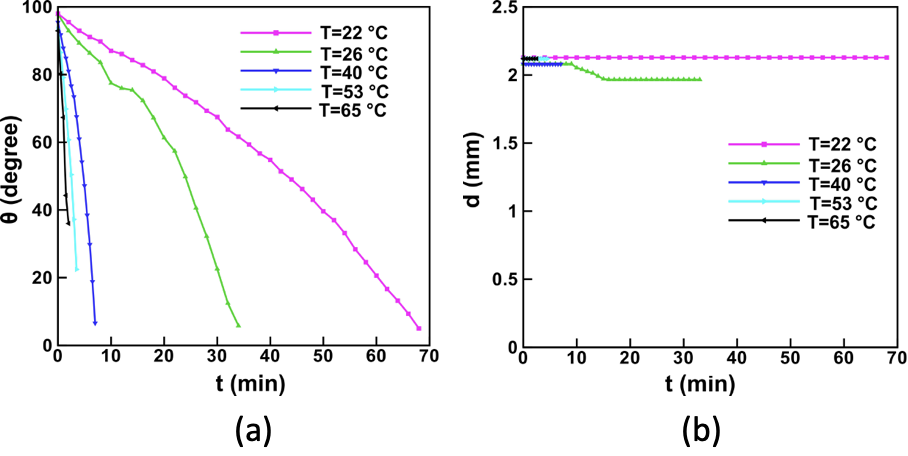}
    \caption{Variation of (a) contact angle and (b) contact diameter with evaporation time for different substrate temperatures.}
    \label{fig:theta}
\end{figure*}

Figs.~\ref{fig:theta}(a) and (b) show the variation of contact angle and contact diameter with evaporation time at different substrate temperatures. No significant variation in the initial contact angle or the initial contact diameter is observed with substrate temperature. Minor deviations may arise due to experimental uncertainties and slight variations in the deposited droplet volume. For the non-heated case, contact angle decreases linearly during the initial pinned contact-line stage for about 10 minutes. Beyond this period, the droplet diameter decreases while maintaining an approximately constant contact angle of $\theta \approx 80^\circ$ for the next 10 minutes. Subsequently, the evaporation proceeds again under pinned contact-line conditions for the remaining evaporation duration. When the substrate is cooled to $T_s = 22^\circ$C, the evaporation rate becomes significantly slower, and the droplet requires around 69 minutes, which is more than two times of non-heated case, to evaporate completely. In this case, the contact line remains pinned throughout the evaporation process, while the contact angle decreases linearly with time. In contrast, when the substrate temperature exceeds the ambient temperature ($T_s > 26^\circ$C), the total evaporation time decreases with a faster reduction in the contact angle at pinned contact line.

\begin{figure*}[htb!]
    \centering
    \includegraphics[width=1\textwidth]{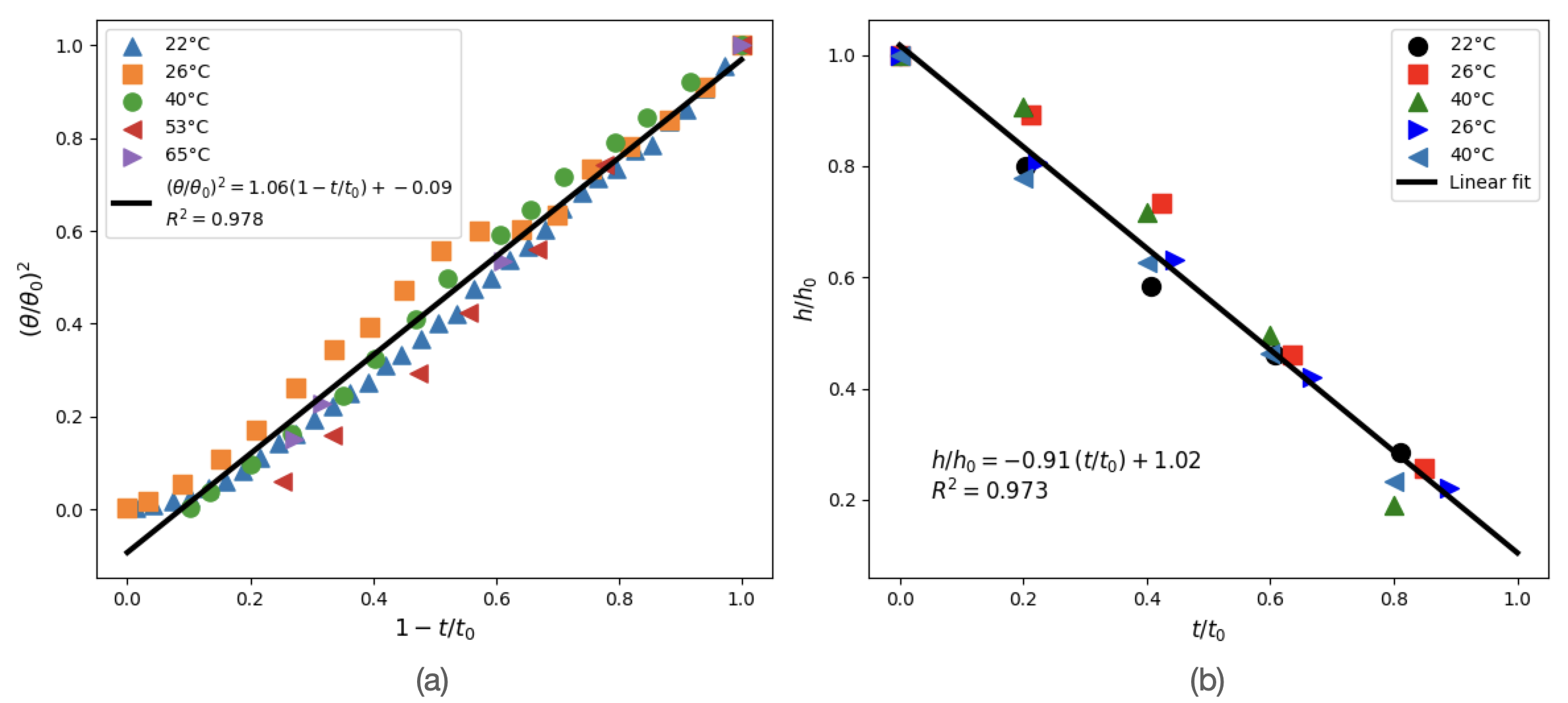}
    \caption{(a) Master curve showing the collapse of the squared normalized contact angle $\left(\theta/\theta_0\right)^2$ as a function of normalized time $\left(1-t/t_0\right)$ for different substrate temperatures. (b) Master curve showing the collapse of normalized droplet height $\left(h/h_0\right)$ with normalized time $\left(t/t_0\right)$ for different substrate temperatures}
    \label{fig:master_angle}
\end{figure*}

To obtain a universal relationship between the contact angle and evaporation time, independent of substrate temperature, the square of the normalized contact angle $\Big(\frac{\theta}{\theta_0}\Big)^2$ is plotted against the normalized evaporation time $\Big(1-\frac{t}{t_0}\Big)$ for all substrate temperatures, as shown in Fig.~\ref{fig:master_angle}(a). A good collapse of the data is observed for all values of $T_s$, indicating a universal evaporation behavior.

A linear fit to the collapsed data yields the following relationship:

\begin{equation}
\Bigg(\frac{\theta}{\theta_0}\Bigg)^2 =
1.06\Bigg(1-\frac{t}{t_0}\Bigg) - 0.09
\end{equation}

This relation can be expressed in the scaling form by ignoring the constant which is very small;

\begin{equation}
\Bigg(\frac{\theta}{\theta_0}\Bigg)^2
\sim
\Bigg(1-\frac{t}{t_0}\Bigg)
\label{master_angle}
\end{equation}

The coefficient of determination is $R^2 = 0.978$, indicating good agreement between the experimental data and the proposed scaling relation. Minor deviations from the scaling behavior may arise due to experimental uncertainties and slight variations in droplet deposition. The scaling relation given by Eq~\ref{master_angle} is consistent with the diffusion-limited evaporation model developed by Yurii O. Popov (2005) \cite{Popov2005} for pinned droplets, thereby supporting the validity of the experimental observations reported in this study
In addition to the contact angle evolution, we also measured the variation of droplet height during evaporation. The nondimensional droplet height $\left(\frac{h}{h_0}\right)$ is plotted as a function of nondimensional time $\left(\frac{t}{t_0}\right)$ for all substrate temperatures. As shown in Fig.~\ref{fig:master_angle}(b), the experimental data collapse well onto a single curve, indicating a universal evaporation behavior independent of substrate temperature.

The collapsed data exhibit an approximately linear decrease of droplet height with time. A linear fit to the data yields the following relation between the normalized droplet height $h$ and evaporation time $t$:

\begin{equation}
\frac{h}{h_0} = -0.91\Bigg(\frac{t}{t_0}\Bigg) + 1.02
\end{equation}

The minor deviation is possibly the contribution of the receding of the contact line at $T_s$ = 26 $^\circ$C and uncertainty in measurements. Therefore, this relationship can be approximated in the following scaling form

\begin{equation}
\frac{h}{h_0} \sim 1 - \Bigg(\frac{t}{t_0}\Bigg)
\label{master_height}
\end{equation}

By combining Eq.~\ref{master_angle} and Eq.~\ref{master_height}, a direct relationship between the droplet contact angle and droplet height can be obtained. The resulting scaling relation is given as

\begin{equation}
\Bigg(\frac{\theta}{\theta_0}\Bigg)^2 \sim \frac{h}{h_0}
\label{angle_height}
\end{equation}

The combined scaling relations  $\left(\theta/\theta_0\right)^2$ and the normalized droplet height $\left(h/h_0\right)$ exhibit linear scaling with the normalized evaporation time. The good collapse of experimental data for all substrate temperatures indicates that the droplet evaporation follows a self-similar evolution despite the variation in thermal conditions. Furthermore, the derived relation $\left(\theta/\theta_0\right)^2 \sim h/h_0$ suggests that the geometrical evolution of the droplet remains coupled during evaporation and can be described using a universal scaling framework. These observations indicate that the overall evaporation dynamics are governed primarily by the geometric evolution of the droplet, while the effect of substrate temperature mainly alters the evaporation timescale without modifying the underlying scaling behavior.
The volume and surface area of a sessile droplet approximated as a spherical cap is given by
\begin{equation}
V = \frac{\pi h}{6} \left( 3R^2 + h^2 \right)~~~~~~~~  A_s = \pi~(R^2+h^2)^{1/2}
\end{equation}
where $h$ is the droplet height and $R$ is the contact radius. $A_s$ is the surface area along the liquid-air interface of the droplet.

During the evaporation process at $T_s = 26^\circ$C, the contact radius remains approximately constant with only minor variations. Therefore, the volume can be non-dimensionalised using $R^3$. Introducing the dimensionless quantities
\begin{equation}
V^* = \frac{V}{R^3}, \qquad h^* = \frac{h}{R},
\end{equation}
the above relation becomes
\begin{equation}
V^* \sim \dfrac{\pi}{6}h^* \left(3 + h^{*2}\right).
\label{eq:scaled_volume}
\end{equation}

The scaled volume can be expressed as

\begin{equation}
V \sim \frac{\pi}{6}R^2~h_0 \left(1-\frac{t}{t_0}\right)\left[3+\frac{h_0^2}{R^2}\left(1-\frac{t}{t_0}\right)^2\right],
\end{equation}
where $V_0$ denotes the initial droplet volume .

This scaled relation does not provide the exact droplet volume but captures the trend of the volume decrease with evaporation time $t$.

As evaporation proceeds, the droplet height decreases. In the later stages of evaporation when $h^2 \ll 3R^2$, the quadratic term becomes negligible and the volume scaling reduces to
\begin{equation}
V \sim \frac{\pi}{2}  R^2 h_0~\left(1-\frac{t}{t_0}\right).
\end{equation}

\begin{figure}[htb!]
    \centering
    \includegraphics[width=1\linewidth]{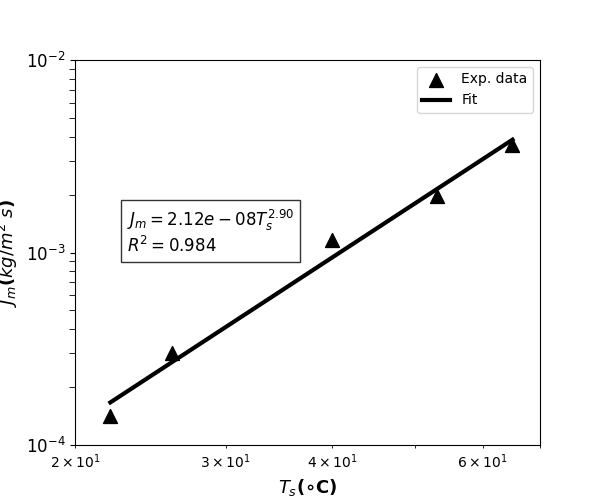}
    \caption{Experimentally measured average evaporative flux $J$ as a function of the substrate temperature $T_s$ along with the power-law fitting.}
    \label{fig:J}
\end{figure}

Fig.~\ref{fig:J} shows the variation of the average evaporative flux ($J_m$) as a function of the substrate temperature ($T_s$). The instantaneous evaporative flux $J$ is calculated as

\begin{equation}
    J = \frac{1}{A_s}\frac{dV}{dt}
    \label{eq:J}
\end{equation}

where $A_s$ is the surface area of the droplet and $dV/dt$ represents the volumetric evaporation rate. For each substrate temperature, the $J$ is calculated over the entire evaporation duration and then averaged over time to obtain the mean evaporative flux.

It is observed that the evaporative flux increases with increasing substrate temperature. The increase in $J_m$ with $T_s$ follows a nonlinear trend, indicating enhanced evaporation at higher temperatures. The experimental data are well described by a power-law relation

\begin{equation}
    J_m \sim 2.12 \times 10^{-8} \, T_s^{2.9}
\end{equation}

with a coefficient of determination $R^2 = 0.984$, indicating excellent agreement between the fitted model and the experimental measurements. This increase in $J_m$ is attributed to the increase in the diffusion coefficient and the enhanced vapor concentration difference at the liquid--air interface due to the rise in the interfacial temperature \cite{Hu2002,Bhardwaj2009}.

\subsection{Deposition Pattern}

\begin{figure*}[htb!]
    \centering
    \includegraphics[width=1\textwidth]{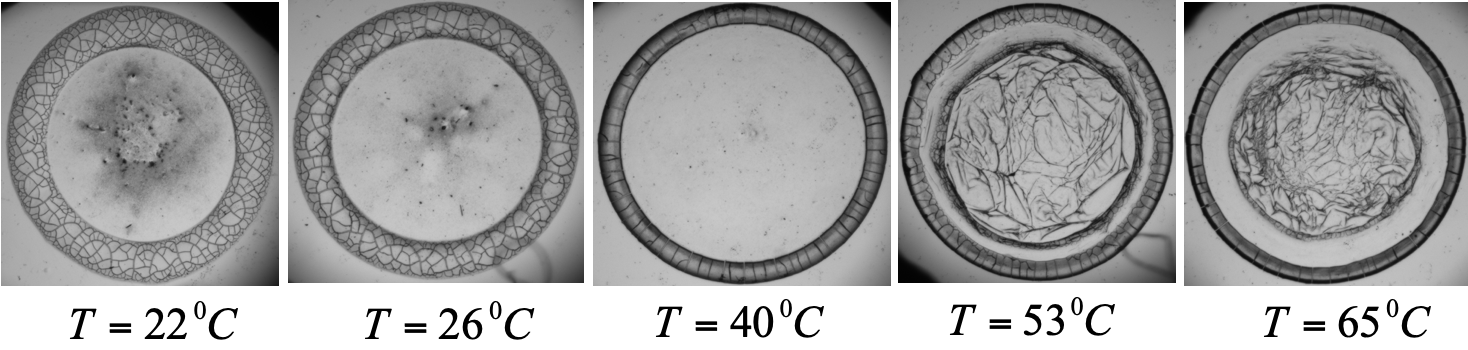}
    \caption{Deposition patterns formed after evaporation of Al$_2$O$_3$ nanofluid droplets at different substrate temperatures.}
    \label{fig:deposition}
\end{figure*}

Fig.~\ref{fig:deposition} shows the deposition patterns obtained after complete evaporation of Al$_2$O$_3$ nanofluid droplets for substrate temperatures $T_s$ = 22$^\circ$C, 26$^\circ$C, 40$^\circ$C, 53$^\circ$C, and 65$^\circ$C. When the droplet evaporates on a non-heated substrate ($T_s$ = 26$^\circ$C), the nanoparticles predominantly accumulate near the contact line, resulting in the classical coffee-ring deposition pattern. The peripheral deposition reveals a
framework of interconnected irregular polygonal tessellation network structures. This type of deposition appears to be unique and, to the best of our knowledge, has not been reported in the existing literature. While most previous studies typically report ring-like (coffee-ring) or relatively uniform deposits, the present polygon-like interparticle network suggests that the final deposit retains the memory of a complex internal flow structure during slow evaporation rate \cite{Deegan1997,Deegan2000,picknett1977evaporation,Dash2014}.

In contrast, when the droplet evaporates on a cooled substrate ($T_s$ = 22$^\circ$C), a thicker ring is observed along with partial particle deposition in the central region, while the polygon-like interparticle network structure is still preserved within the deposit. The width of the coffee ring is increased.

At an elevated substrate temperature of $T_s$ = 40$^\circ$C, a well-defined and clean coffee-ring pattern is observed without any noticeable central deposition. However, the ring width decreases compared to the lower temperature cases due to the enhanced evaporation rate as shown in Fig \ref{fig:J}, which increase the outward capillary transport of particles toward the contact line during late stage of evaporation, when $\frac{h}{R} \ll 1$\cite{saroj2021magnetophoretic}. A radial cracks formation take within the deposit film. The substrate temperature of $T_s$ = 40$^\circ$C can therefore be considered as a critical temperature, beyond which the deposition behavior begins to transition. At $T_s$ = 53$^\circ$C, the deposition pattern transforms from a single coffee-ring structure to a dual-ring configuration accompanied by partial central deposition, as shown in Fig.~\ref{fig:deposition}. With a further increase in substrate temperature to $T_s$ = 65$^\circ$C, the central deposition becomes more pronounced and the separation distance between the primary and secondary rings increases significantly.

\begin{figure}[htb!]
    \centering
    \includegraphics[width=0.5\textwidth]{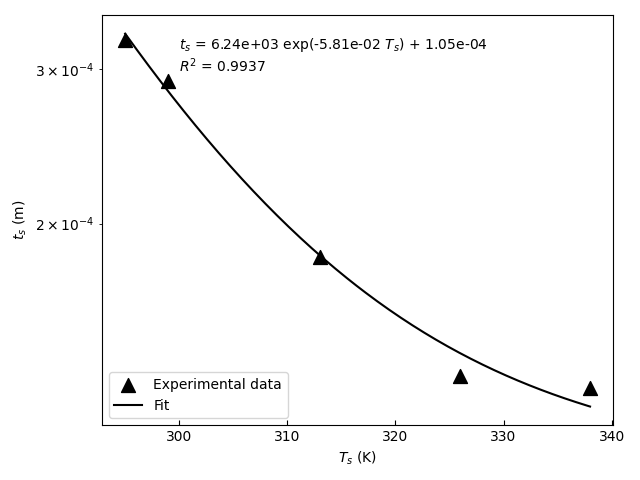}
    \caption{Variation of ring width $t_s$ as a function of substrate temperature $T_s$.}
    \label{fig:thick}
\end{figure}

Figure~\ref{fig:thick} shows the variation of the ring width $t_s$ as a function of substrate temperature $T_s$. The experimental data exhibit an exponential decrease in $t_s$ with increasing substrate temperature. The exponential fitting yields a coefficient of determination $R^2 = 0.994$, indicating excellent agreement with the experimental measurements.


\begin{figure*}[htb!]
    \includegraphics[width=\textwidth]{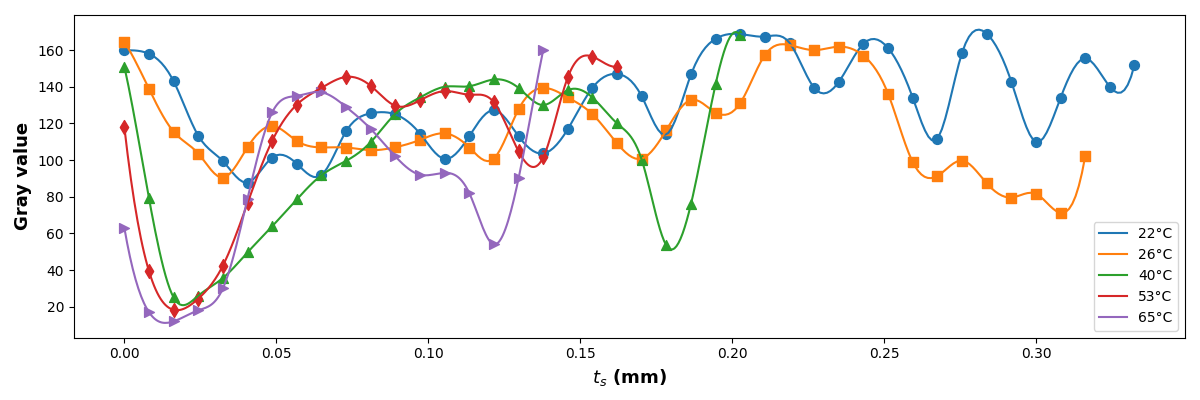}
    \caption{Variation of gray value along $t_s$ for different substrate temperature, $T_s$.}
    \label{fig:GR}
\end{figure*}

Fig.~\ref{fig:GR} shows the variation of gray value along $t_s$, of the deposition pattern. The gray value quantifies the particle deposition density: a higher gray value corresponds to a lower particle density, and vice versa. The higher particle density at the outer edge of the ring appears darker (black) in the magnified images shown in Fig.~\ref{fig:deposition}, particularly for heated substrates, and corresponds to the minimum gray values observed in Fig.~\ref{fig:GR} for $T_s > 26,^\circ$C. The interconnected particle framework irregular polygonal tessellation network structure for substrate temperatures $T_s \leq 26,^\circ$C and the corresponding gray value exhibits a gradual increase and decrease along the $t_s$, indicating non-uniform particle packing. In contrast, for $T_s > 26,^\circ$C, the gray value remains approximately uniform across the $t_s$, suggesting a more homogeneous particle distribution.

\subsection{Temperature distribution}
\subsubsection{Interface temperature }

\begin{figure*}[htb!]
    \centering
    \includegraphics[width=1\textwidth]{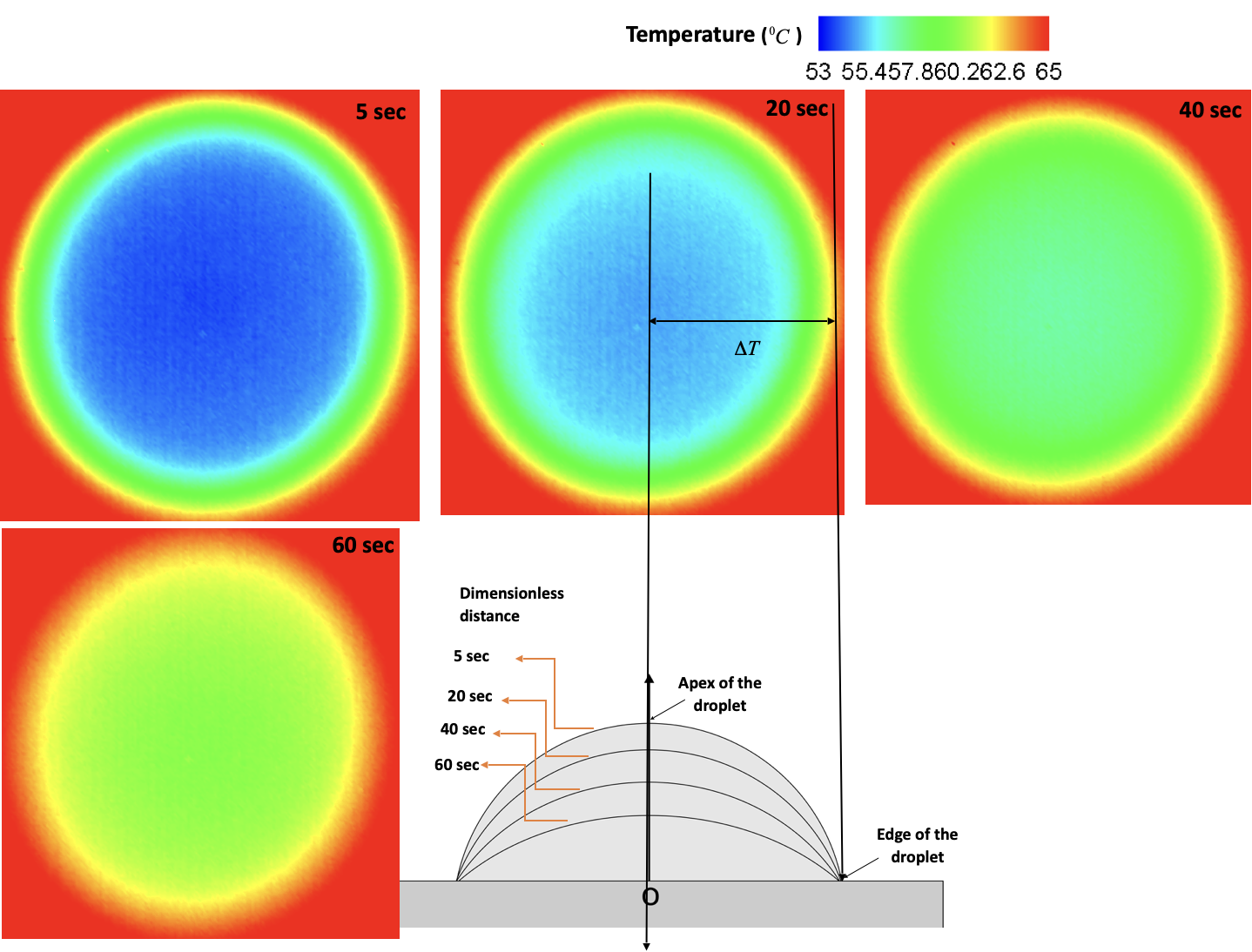}
    \caption{Infrared thermography images showing the temperature distribution over the droplet interface at $T_s$ = 65$^\circ$C for different evaporation times.}
    \label{fig:surf_temp}
\end{figure*}

Fig.~\ref{fig:surf_temp} shows the recorded infrared (IR) thermography images of the temperature distribution along the liquid–air interface of the droplet for $T_s$ = 65$^\circ$C at different stages of evaporation. The temperature measurements were performed during the initial stage of evaporation, as indicated in the schematic shown in Fig.~\ref{fig:surf_temp}. 

\begin{figure*}[htb!]
    \centering
    \includegraphics[width=1\textwidth]{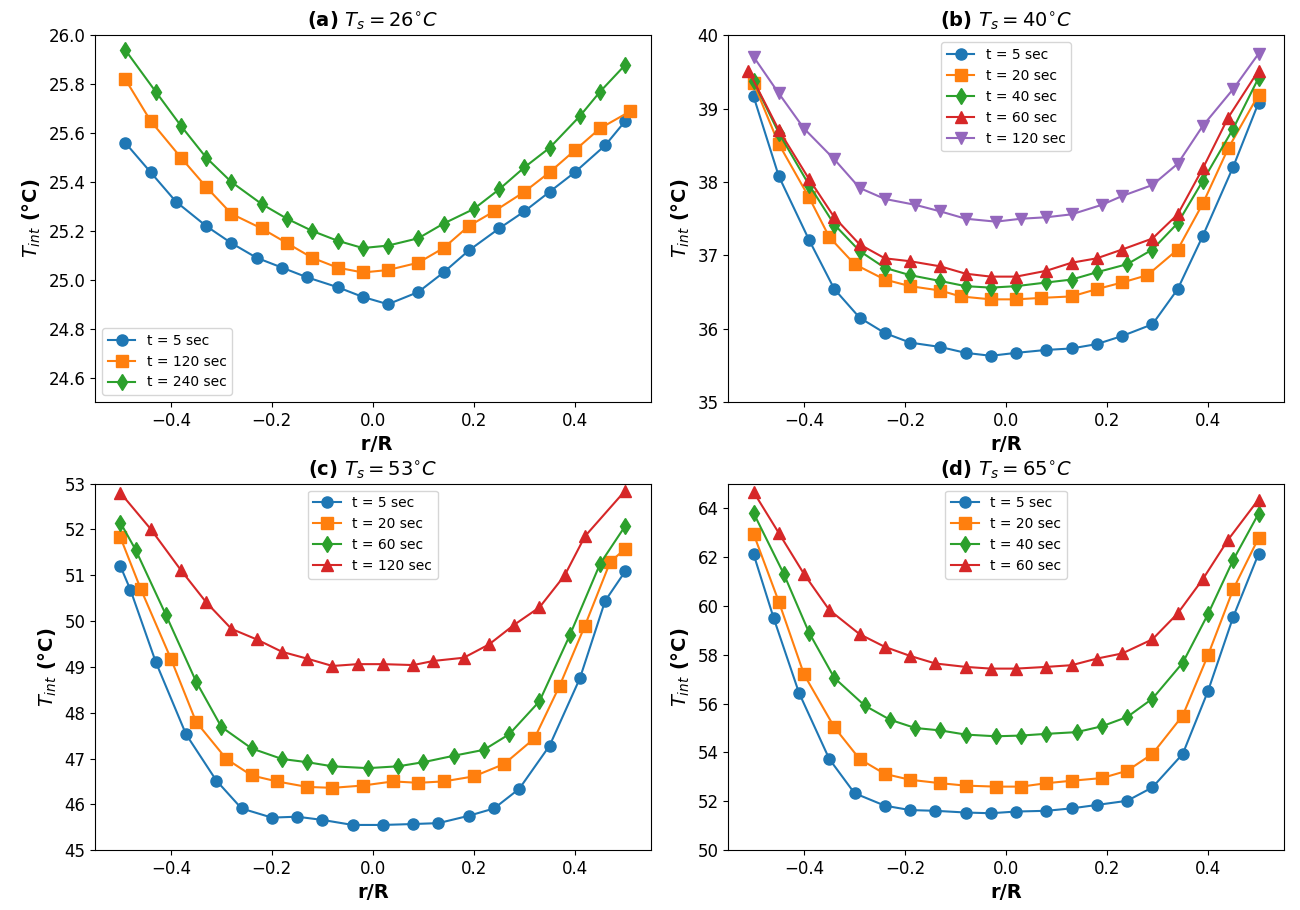}
    \caption{Temperature variation along the droplet interface as a function of normalized radial position $(r/R)$ for different substrate temperatures and evaporation times: (a) $T_s$ = 26$^\circ$C, (b) $T_s$ = 40$^\circ$C, (c) $T_s$ = 53 $^\circ$C and (a) $T_s$ = 65$^\circ$C.}
    \label{fig:temp}
\end{figure*}

Fig.~\ref{fig:temp} shows the variation of the interfacial temperature along the normalized radial position $(r/R)$ of the droplet for different substrate temperatures and evaporation times. The temperature distribution along the droplet interface is symmetric about the radial center, confirming the uniform heating condition of the substrate. This symmetry also explains the symmetric deposition patterns obtained after complete evaporation.

It is clearly observed that the interfacial temperature is highest near the contact line and decreases towards the droplet apex. Furthermore, the interface temperature $T_{int}$ increases with evaporation time for all substrate temperatures. The temperature difference between the edge and apex, defined as $\Delta T = T_{edge} - T_{apex}$, gradually decreases with evaporation time for heated substrates. However, in the non-heated case ($T_s$ = 26$^\circ$C), $\Delta T$ remains nearly constant throughout the evaporation process.

\begin{figure}[htb!]
    \centering
    \includegraphics[width=0.5\textwidth]{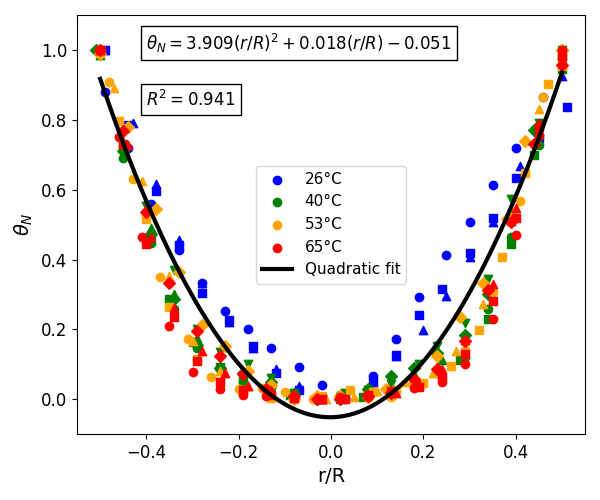}
    \caption{Collapse of normalized interfacial temperature ($\theta_N $) profiles showing a universal quadratic dependence on normalized radial position $(r/R)$.}
    \label{fig:universal}
\end{figure}

The universal interfacial temperature profile is obtained by plotting the normalized temperature, $\theta_N = \frac{T_{int}-T_{apex}}{T_{edge}-T_{apex}}$
against the normalized radial distance $\frac{r}{R}$. All the experimental data collapse onto a single quadratic curve, as shown in Fig.~\ref{fig:universal}. The resulting quadratic fit provides a universal interfacial temperature profile.


\begin{equation}
\theta_N = 3.91 \Big(\frac{r}{R}\Big)^2 + 0.018 \Big(\frac{r}{R}\Big)-0.051
\end{equation}

where $T_{edge}$ and $T_{apex}$ represent the temperatures at the contact line and apex of the droplet, respectively. The coefficient of determination $R^2 = 0.941$ indicates reasonably good agreement with the experimental data. Minor deviations may arise due to measurement uncertainties.

The above expression is further simplified for the dominating quadratic term,

\begin{equation}
\theta_N \sim 3.91 \Big(\frac{r}{R}\Big)^2
\end{equation}

Therefore, the interfacial temperature distribution along the droplet interface can be expressed as

\begin{equation}
T_{int} = T_{apex} + 3.91 (T_{edge}-T_{apex}) \Big(\frac{r}{R}\Big)^2
\end{equation}

This quadratic dependence suggests that heat transfer along the droplet interface is predominantly conduction-driven, with additional contributions from thermo-capillary (Marangoni) convection.

\begin{figure*}[htb!]
    \centering
    \includegraphics[width=1\textwidth]{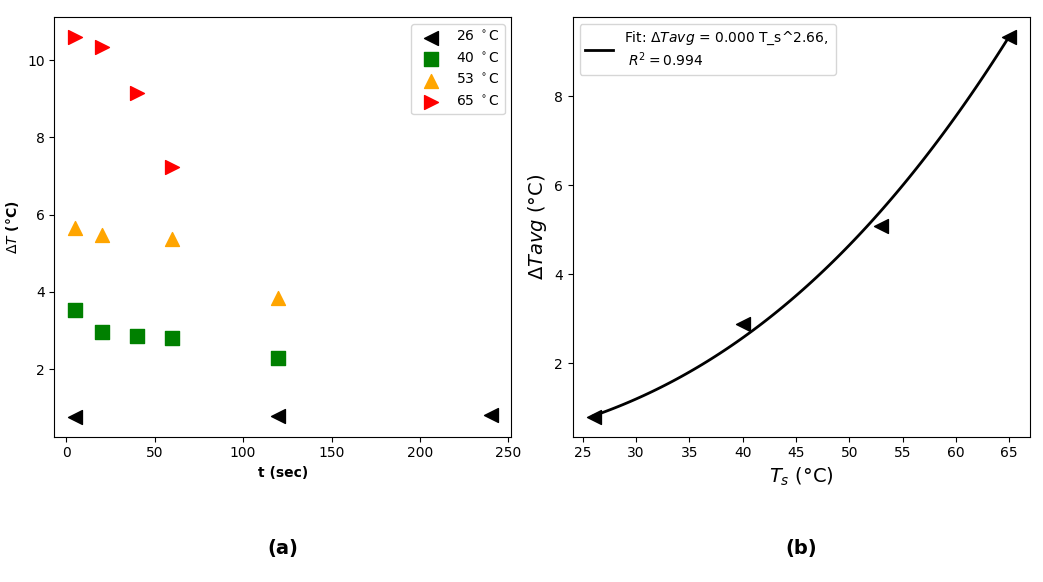}
    \caption{(a) Variation of temperature difference $\Delta T = T_{edge}-T_{apex}$ with evaporation time for different $T_s$, and (b) average $\Delta T$ over time as a function of $T_s$. The red symbols and black solid line represent the experimental data and power law fitting, respectively.}
    \label{fig:delT}
\end{figure*}

Fig.~\ref{fig:delT} (a) shows the variation of the temperature difference $\Delta T$ as a function of substrate temperature $T_s$ for different evaporation times, along with the time-averaged values in Fig.~\ref{fig:delT} (b). The temperature difference generally increases with evaporation time, and the highest $\Delta T$ is observed for $T_s$ = 65$^\circ$C.

To obtain a scaling relation between $\Delta T$ and $T_s$, the time-averaged temperature difference $\Delta T_{avg}$ was plotted against $T_s$, as indicated by the red symbols in Fig.~\ref{fig:delT} (b). The experimental data follow a power-law scaling given by

\begin{equation}
\Delta T_{avg} \sim 0.000143\, T_s^{2.66}
\label{delT}
\end{equation}

with a coefficient of determination $R^2 = 0.994$, indicating an excellent fit to the experimental data. This scaling demonstrates a strong nonlinear dependence of the interfacial temperature difference on the substrate temperature.

\subsubsection{Temperature distribution inside droplet}

\begin{figure}
    \centering
    \includegraphics[width=1\linewidth]{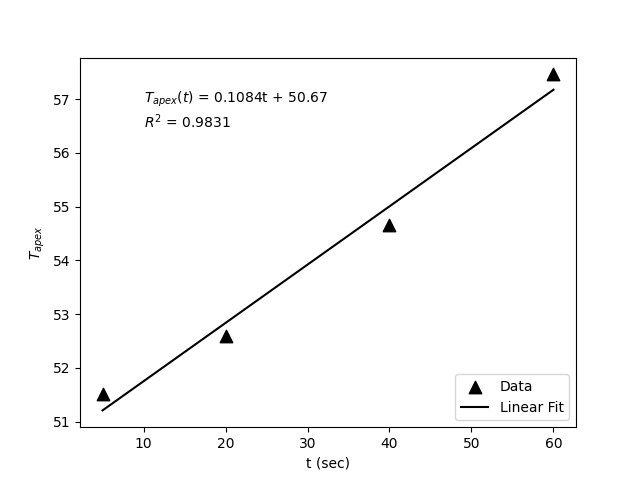}
    \caption{Variation of $T_{apex}$ as a function of evaporation time $t$ for $T_s$ = 65 $^\circ$C along with the linear fit. Symbols and solid line denote the experimental data and linear fit, respectively.}
    \label{fig:Tapex}
\end{figure}

The apex temperature, $T_{\text{apex}}$, is extracted at different time instants from Fig.~\ref{fig:temp} for $T_s = 65^\circ$C. The variation of $T_{\text{apex}}$ with time is shown in Fig.~\ref{fig:Tapex}, which exhibits a linear increase with time. A linear fit to the data yields:

\begin{equation}
T_{\text{apex}}(t)= 0.1084\,t + 50.67 \qquad (R^2 = 0.983)
\label{Tapex}
\end{equation}

This indicates that the apex temperature increases at an approximately constant rate of 
$\frac{dT_{\text{apex}}}{dt} \approx 0.1084~^\circ$C/s.

Owing to the small height of the droplet ($h_0 = 1.3$ mm), heat transfer within the droplet can be approximated as one-dimensional along the vertical direction. The governing energy equation is given by:

\begin{equation}
k_{nf} \frac{d^2 T}{dh^2} + q_v = \rho_{nf} c_{nf} \frac{dT}{dt}
\end{equation}

Assuming that the temporal temperature variation within the droplet follows that of the apex, i.e., 
$\frac{dT}{dt} \approx \frac{dT_{\text{apex}}}{dt}$, and that the volumetric heat generation balances the rate of energy increase, such that 
$q_v = \rho_{nf} c_{nf} \frac{dT}{dt}$, the governing equation reduces to:

\begin{equation}
\frac{d^2 T}{dh^2} = 0
\end{equation}

Applying the boundary conditions:

\begin{equation}
T(0,t) = T_s, \qquad T(h_0,t) = T_{\text{apex}}(t)
\end{equation}

the temperature distribution inside the droplet is obtained as:

\begin{equation}
T(h,t) = T_s + \frac{T_{\text{apex}}(t) - T_s}{h_0}\, h
\label{finalT}
\end{equation}

This result indicates that the temperature varies linearly along the droplet height, while the apex temperature increases linearly with time. The resulting temperature gradient plays a key role in the development of thermo-capillary flow.

\subsection{Flow visualization}
\begin{figure*}[htb!]
    \centering
    \includegraphics[width=1\textwidth]{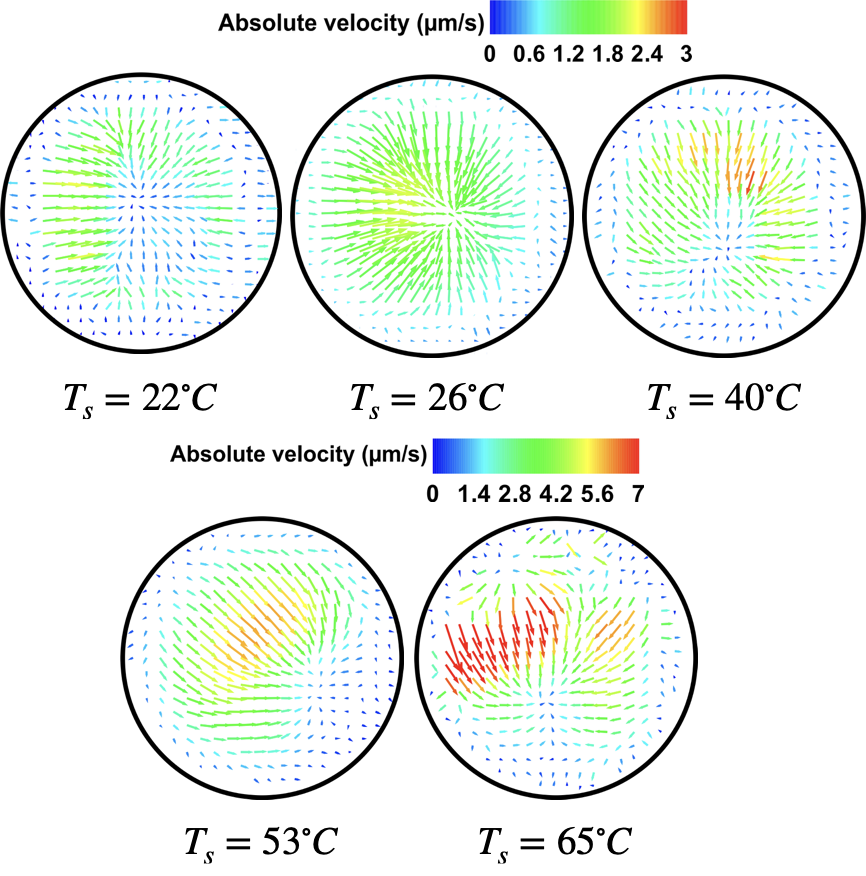}
    \caption{Velocity field distribution inside an evaporating droplet obtained from $\mu$-PIV measurements at different substrate temperatures. The internal flow exhibits a circulatory pattern that strengthens with increasing substrate temperature due to enhanced evaporation and thermocapillary effects. The corresponding visualization movies are provided in supplementary material.}
    \label{fig:piv}
\end{figure*}

To obtain the velocity distribution inside the droplet, $\mu$-PIV measurements were carried out for all substrate temperatures $T_s$. Fig.~\ref{fig:piv} illustrates the velocity field distribution measured at 50~$\mu$m above the substrate surface at different substrate temperatures. The vector field is reported for 10\% of the total evaporation time.

The flow inside the droplet moves from the edge region towards the central region at $T_s$ = 22 and 26 $^\circ$C. With close observation  it is seen that the particles near the contact line move towards the peripheral region with relatively low velocity. The velocity near the contact line region is relatively small and increases towards the center, followed by nearly zero velocity at the center for $T_s$ = 22 $^\circ$C at 50~$\mu$m above the substrate surface . The magnitude of velocity increases with increase of the $T_s$ from 22 $^\circ$C to 26 $^\circ$C.   The circulatory motion persists throughout most of the evaporation process. However, during the later stages of evaporation, when $\frac{h}{R} \ll 1$, the flow gradually transitions toward a radial flow directed to the contact line, as evidenced by observations reported in our recent study \cite{saroj2021magnetophoretic}.

It is also observed that the magnitude of the internal velocity increases with increasing substrate temperature. A slight asymmetry in the velocity field is observed at moderate substrate temperature ($T_s$ = 40$^\circ$C), as shown in Fig.~\ref{fig:piv}. With further increase in substrate temperature to $T_s$ = 53$^\circ$C, the flow strength increases significantly and a toroidal vortex structure develops inside the droplet along with the asymmetry in the flow pattern. Readers are referred to the supplementary movie 1 provided for a clearer visualization of the internal flow dynamics.

The increase in velocity with substrate temperature is attributed to the higher evaporation rate and the enhanced thermo-capillary (Marangoni) convection induced by the interfacial temperature gradients.

The enhancement of the internal circulation with increasing substrate temperature suggests a stronger thermo-capillary driving force, which can be associated with an increase in the Marangoni number.
The maximum velocity $U_{max}$ obtained from Fig.~\ref{fig:piv} is extracted and plotted as a function of the substrate temperature $T_s$. The $U_{max} \propto T_s$, indicating that the internal flow velocity within the droplet increases almost linearly with increasing substrate temperatures. Also, upon close observation of Movie~1 as well as Fig.~\ref{fig:piv} for  $T_s = 53$ and $65^\circ$C, an asymmetry in the flow pattern along with the  appearance of a few vortices is observed, indicating the onset of Marangoni instability, which will be discussed in a later section.

\begin{figure}[htb!]
    \centering
    \includegraphics[width=0.5\textwidth]{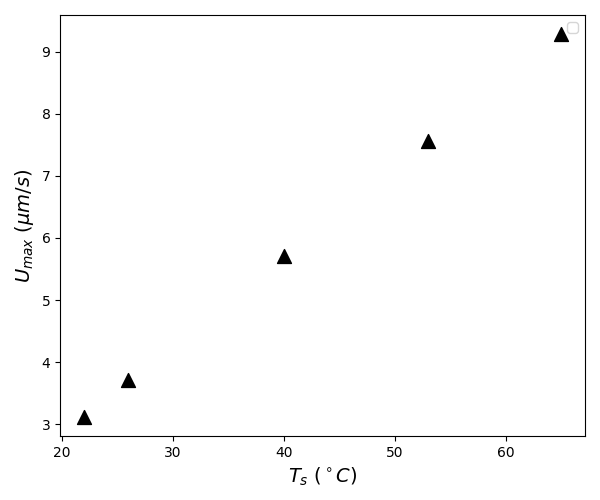}
    \caption{Variation of the maximum internal velocity, $U_{max}$, extracted from Fig. \ref{fig:piv}, with substrate temperature $T_s$. The maximum velocity increases linearly with increase of $T_s$.}
    \label{fig:velmax}
\end{figure}


\subsection{Evaporative cooling}

The evaporative cooling resulting from droplet evaporation can be quantified by the temperature difference between the substrate and the liquid–air interface, i.e., $T_s - T_{int}$. Since $T_s - T_{int}$ decreases from the apex towards the contact line, the evaporative cooling is expected to be strongest near the contact line region where higher evaporation occurs. 

\begin{figure*}[htb!]
    \centering
    \includegraphics[width=1\linewidth]{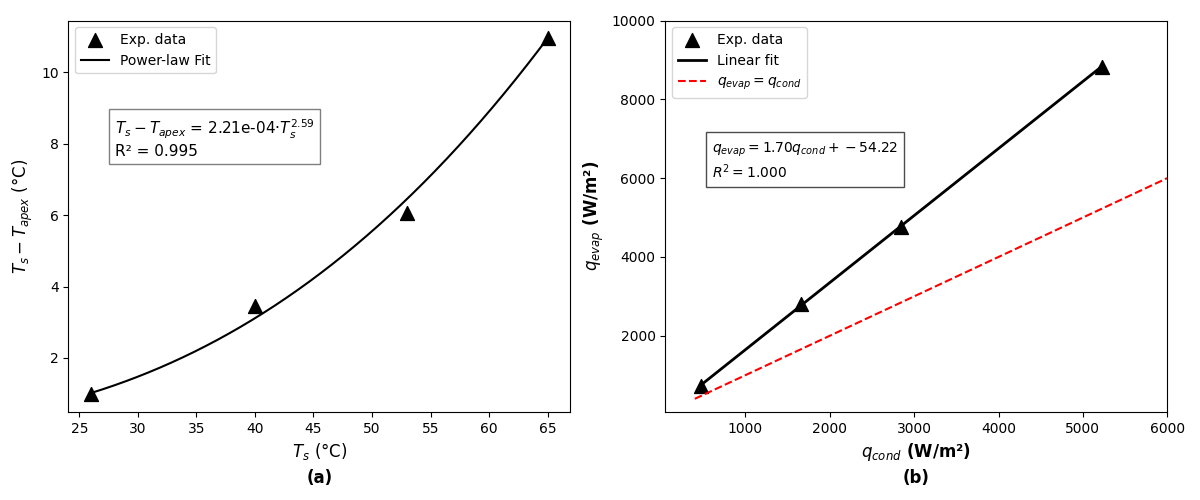}
    \caption{(a) Variation of the temperature difference between the substrate and the droplet apex, $T_s - T_{apex}$, as a function of substrate temperature. (b) Comparison between evaporative heat removal and conductive heat flux supplied from the substrate. The symbols represent experimental data and the solid line denotes the linear fit. The dotted red line represents the condition $q_{evap} = q_{cond}$. $q_{cond}$ is calculated using the apex temperature difference, while $q_{evap}$ is obtained from the global average evaporation flux.}
    \label{fig:cooling}
\end{figure*}

Fig.~\ref{fig:cooling}(a) shows the variation of the averaged temperature difference between the substrate and the droplet apex, $T_s - T_{apex}$, as a function of substrate temperature $T_s$. The results show a nonlinear increase in the $ \Delta T_s$ with increasing substrate temperature. The scaled power-law fitting predicts the following relation

\begin{equation}
    T_s - T_{apex} \sim 2.21 \times 10^{-4} ~T_s^{2.59}~~~~~~(R^2=0.995)
\end{equation}
The variation of $\Delta T_s$ follows the trend $\Delta T \propto T_s^{2.66}$. 
Therefore, evaporative cooling generates a temperature difference 
$\Delta T$ along the droplet interface. The heat flux conducted through the droplet, assuming conduction-dominated heat transfer, can be expressed as

\begin{equation}
    q_{cond} = \frac{k_{nf}}{h_0}\,(T_s - T_{int})
\end{equation}

The cooling of the droplet occurs due to the absorption of latent heat during evaporation. Therefore, the evaporative heat removal can be expressed as

\begin{equation}
    q_{evap} = J\,h_{fg}
\end{equation}

To quantify the evaporative cooling, a cooling effectiveness parameter is defined as

\begin{equation}
\epsilon = \frac{q_{evap}}{q_{cond}}
\sim
\frac{h_0\,J\,h_{fg}}{k_{nf}\,(T_s - T_{int})}
\end{equation}

Substituting the expression for the interfacial temperature distribution gives

\begin{equation}
\epsilon
\sim
\frac{h_0\,J\,h_{fg}}
{k_{nf}\left[T_s - T_{apex} - 3.91 (T_{edge}-T_{apex})\left(\frac{r}{R}\right)^2\right]}
\end{equation}

The evaporation flux $J$ can be expressed as

\begin{equation}
J \sim 
\frac{\epsilon k_{nf}\left[T_s - T_{apex} - 3.91 (T_{edge}-T_{apex})\left(\frac{r}{R}\right)^2\right]}
{h_0 h_{fg}}
\label{J_T}
\end{equation}

By considering $\Delta T_s \propto T_s^{2.59}$, the evaporation flux scales as

\begin{equation}
J \propto T_s^{2.59}\left(1 - 3.91\left(\frac{r}{R}\right)^2\right).
\end{equation}

This result indicates that the evaporation flux increases strongly with 
substrate temperature and exhibits a quadratic radial dependence along 
the droplet interface.

The cooling effectiveness can be evaluated at the apex, i.e., $T_{int}=T_{apex}$, which yields

\begin{equation}
\epsilon
\sim
\frac{h_0\,J\,h_{fg}}{k_{nf}(T_s - T_{apex})}
\end{equation}

The calculated cooling effectiveness  $\epsilon$ $>$1 for all $T_s$ are presented in Table \ref{Ja}, indicating that the evaporative heat removal exceeds the heat supplied by conduction from the substrate. Also, the $\epsilon$ is approximately equal for heated substrate i.e $T_s$ $>$ 26 $^\circ$C.

Fig.~\ref{fig:cooling}(b) shows the variation between the conductive heat flux $q_{cond}$ and the evaporative heat removal $q_{evap}$. The symbols represent the experimental data and the solid line denotes the linear fit, which yields

\begin{equation}
    q_{evap} = 1.7~q_{cond} - 54.22
\end{equation}

The dotted red line represents the condition $q_{evap} = q_{cond}$. The experimental data lie above this line, indicating that the evaporative heat removal is larger than the heat supplied purely by conduction from the substrate to the droplet apex. This apparent discrepancy arises due to the simplifying assumptions used in the estimation of the conductive heat flux. In the present analysis, $q_{cond}$ is evaluated using the temperature difference between the substrate and the apex of the droplet. However, the droplet hight decreases toward the contact line, resulting in a shorter conduction path and therefore stronger conductive heat transfer in the edge region. Consequently, a significant portion of the heat supplied to the droplet likely originates from the contact line region, which is not captured when conduction is estimated only at the apex.

Furthermore, the evaporative heat flux $q_{evap}$ is calculated using the global average evaporation flux $J$, which accounts for evaporation over the entire droplet surface. Therefore, the comparison between the apex-based conductive heat flux and the global evaporative heat flux leads to an apparent imbalance, yielding $\epsilon \approx 1.68$ for heated substrate . In addition, internal thermo-capillary (Marangoni) convection may redistribute heat within the droplet, further contributing to the energy transport required for evaporation. The change in phase during evaporation enhances the cooling.

Additionally, the Jakob number ($Ja$), which compares the evaporative cooling to the heat conducted through the droplet to the interface, is defined as:

\begin{equation}
Ja = C_{nf} \frac{T_s - T_{apex}}{h_{fg}}
\end{equation}

Using the experimentally obtained scaling for the temperature difference, the Jakob number can be expressed as:

\begin{equation}
Ja \sim \frac{2.21 \times 10^{-4} \, T_s^{2.59} \, C_{nf}}{h_{fg}}
\label{Ja1}
\end{equation}

The calculated values of $Ja$ for all substrate temperatures are listed in Table~\ref{Ja}. Here, $h_{fg} = 2.434 \times 10^{6} \,\mathrm{J\,kg^{-1}}$. It is observed that $Ja \ll 1$ for all $T_s$, indicating that evaporative cooling is significantly higher than the heat conducted through the droplet.
\begin{table}[htb!]
\centering
\caption{Variation of cooling effectiveness and Jakob number for different substrate temperature}
\label{Ja}
\begin{tabular}{|c|c|c|}
\hline
$T_{s}$ ($^\circ$C)&$\epsilon$ & $Ja$ \\
\hline
26 & 1.55&$1.7 \times 10^{-3}$ \\
40 & 1.69&$5.3 \times 10^{-3}$ \\
53 & 1.67&$1.1 \times 10^{-2}$ \\
65 & 1.69&$1.9 \times 10^{-2}$ \\
\hline

\end{tabular}

\end{table}
Therefore, this confirms that the cooling due to latent heat absorption during evaporation dominates over conductive heat transport within the droplet.

The increasing of the Jakob number with increase of $T_s$, indicates the intensification of the evaporation flux at the contact line region resulting in pinning of the contact line compared to that of the evaporation at non-heated substrate.
\section{Discussion}

The capillary length is given by $
l_c = \left(\frac{\sigma}{\rho_{nf} g}\right)^{1/2} = 2.7~\text{mm},
$
$\gg$ $R$. The corresponding Bond number\cite{tonini2023modeling}, $
B_o = \left(\frac{R}{l_c}\right)^2 = 0.158 \ll 1,
$
indicates that surface tension forces dominate over gravitational forces; therefore, the effect of gravity is negligible. The thermal diffusion time scale is estimated as
$
\tau_d = \frac{R^2}{\alpha_{nf}} = 7.8~\text{s}.
$
Since the total evaporation time $t_0 \gg \tau_d$, heat diffusion inside the droplet is much faster than the evaporation process. Therefore, the temperature field within the droplet rapidly reaches a quasi-steady state and the interface temperature becomes nearly steady during evaporation.


\subsection{Rayleigh and Marangoni convection}
The temperature difference between the substrate and the droplet apex, as shown in Fig.~\ref{fig:cooling}(a), together with the approximately linear temperature profile along the vertical direction inside the droplet (as described by Eq.~\ref{finalT}), establishes a temperature gradient within the droplet. This gradient can induce buoyancy-driven flow, commonly referred to as Rayleigh convection. The internal flow may therefore be influenced by buoyancy effects, whose relative importance compared to viscous forces is characterized by the thermal Grashof number, defined as~\cite{batchelor1967fluid,Fox2011}:

\begin{equation}
Gr = \frac{g \beta_{nf} \Delta T_s h_0^{3}}{\nu_{nf}^{2}} = K_{Gr}\,\Delta T_s
\end{equation}

where 
\begin{equation}
K_{Gr} = \frac{g \beta_{nf} h_0^{3}}{\nu_{nf}^{2}} = 5.75~\text{K}^{-1}.
\end{equation}

Furthermore, evaporative cooling leads to an increase in $\Delta T_s$ from the apex toward the contact line, resulting in a temperature gradient along the liquid--air interface of the droplet (see Fig.~\ref{delT}). This interfacial temperature gradient generates a surface tension gradient, with lower surface tension at the warmer contact line and higher surface tension at the cooler apex. Consequently, fluid flows from the warmer region to the colder region, forming a recirculatory motion known as thermo-capillary (Marangoni) flow\cite{Bhardwaj2009,liu2022influence}.

The strength of this thermo-capillary flow is quantified by the Marangoni number, $Ma$, which arises due to temperature-induced surface tension gradients along the interface. It is defined as~\cite{Hu2005Marangoni}:

\begin{equation}
Ma = \frac{R \left(-\frac{d\sigma}{dT}\right) \Delta T}{\mu_{nf} \alpha_{nf}} = K_{ma}\,\Delta T
\end{equation}

where 
\begin{equation}
K_{ma} = \frac{R \left(-\frac{d\sigma}{dT}\right)}{\mu_{nf} \alpha_{nf}} = 1264.6~\text{K}^{-1}.
\end{equation}

Here, $\frac{d\sigma}{dT} = -1.57 \times 10^{-4}~\text{N\,m}^{-1}\text{K}^{-1}$ is the temperature coefficient of surface tension~\cite{Zhu2010}, $\mu_{nf}$ is the dynamic viscosity, $\alpha_{nf}$ is the thermal diffusivity, $\beta_{nf}$ is the volumetric thermal expansion coefficient, and $\Delta T$ is the temperature difference between the contact line and the apex of the droplet, while $\Delta T_s = T_s - T_{int}$.

It is important to note that both viscosity and surface tension decrease with increasing temperature~\cite{Zhu2010}, which further influences the internal flow dynamics.
\begin{figure}[htb!]
    \centering
    \includegraphics[width=1\linewidth]{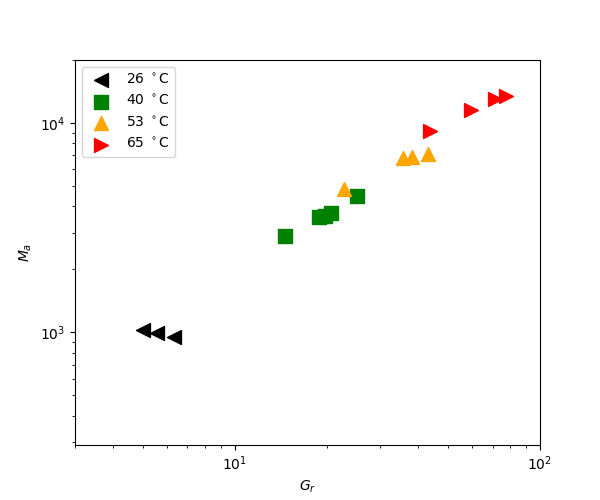}
    \caption{ Regime map of Marangoni number ($Ma$) versus Grashof number ($Gr$) at the initial stage of droplet evaporation for different substrate temperatures ($T_s$).}
    \label{fig:Ma_Gr}
    \end{figure}

 Fig.~\ref{fig:Ma_Gr} shows the variation of the Grashof number ($G_r$) and the Marangoni number ($M_a$), indicating that $M_a \gg G_r$. This suggests that the flow inside the droplet is primarily governed by thermo-capillary (Marangoni) convection, with only a weak contribution from buoyancy-driven flow.

\begin{figure}[htb!]
    \centering
    \includegraphics[width=1\linewidth]{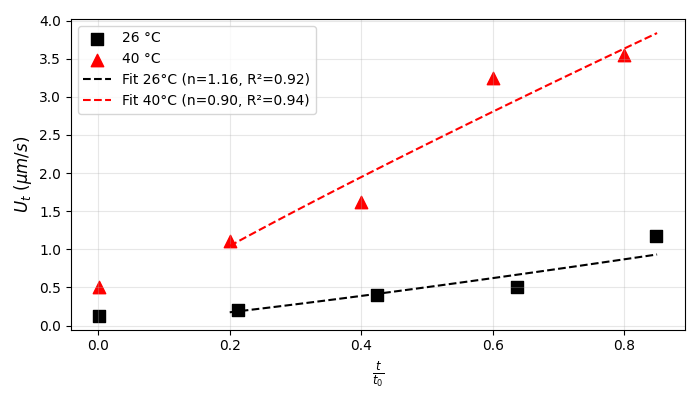}
    \caption{The variation of instantaneous capillary velocity as a function of the normalized evaporation time for $T_s$ = 26  and 40 $^\circ$C.}
    \label{fig:Ut}
\end{figure}

 \begin{figure}[htb!]
    \centering
    \includegraphics[width=1\linewidth]{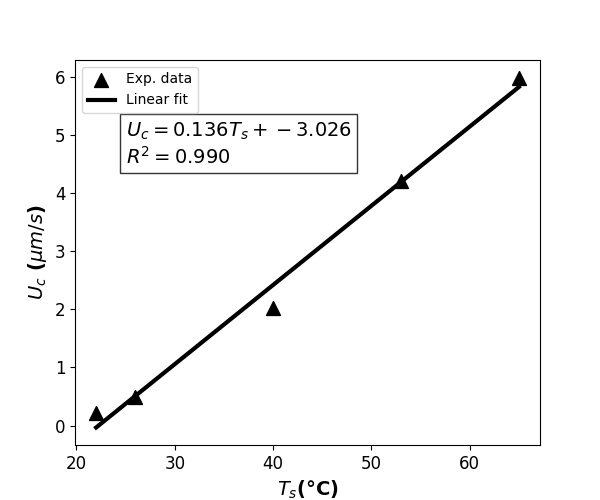}
    \caption{The mean capillary velocity $U_c$ as a function of substrate temperature $T_s$ along with the linear fit.}
    \label{fig:Uc}
\end{figure}


\subsection{Capillary flow}
The temperature gradient along the liquid--air interface (see Fig.~\ref{fig:temp}) leads to non-uniform evaporation, with a higher evaporation rate near the contact line that gradually decreases toward the apex region. To replenish the enhanced liquid loss at the contact line, fluid from the interior of the droplet flows outward toward the contact line. Additionally, the temporal decrease in droplet height further drives this outward flow. This flow mechanism is commonly referred to as capillary flow\cite{Xu2014,Deegan1997,FoxFluidMechanics}. The capillary flow strength is calculated using the expression:

\begin{equation}
    U_t = \frac{J R}{h}
    \label{Ut}
\end{equation}

where $J$ is the evaporative flux, $R$ is the droplet radius, and $h$ is the droplet height. 

Fig.~\ref{fig:Ut} shows the instantaneous capillary velocity calculated using Eq.~\ref{Ut} for $T_s = 26$ and $40\,^\circ\mathrm{C}$ as a function of $t/t_0$. The dotted lines represent the power-law fits, given by
\begin{equation}
U_t \sim \left(\frac{t}{t_0}\right)^n
\end{equation}
with $n = 1.16$ ($R^2 = 0.92$) for $T_s = 26\,^\circ\mathrm{C}$, and $n = 0.9$ ($R^2 = 0.94$) for $T_s = 40\,^\circ\mathrm{C}$.

The results indicate that the capillary velocity increases during the late stage of evaporation. Although the magnitude of the capillary velocity is higher at elevated substrate temperature due to the increased evaporation flux, the lower exponent ($n = 0.9$) suggests a relatively weaker temporal growth compared to the case at $26\,^\circ\mathrm{C}$ ($n = 1.16$). This implies that while evaporation enhances the overall flow intensity, the acceleration of capillary-driven transport is moderated at higher temperature.

Both the increase in evaporative flux $J$ and the progressive decrease of $\frac{h}{R}$ during evaporation lead to a substantial increase in the 
capillary velocity $U_t$ at the late stage of evaporation. Consequently, 
particles are preferentially transported toward the contact line and 
ultimately deposit in the peripheral region even at higher substrate 
temperatures $T_s$. The instantaneous velocity is measured over the evaporation duration for each $T_s$ and then averaged over time to obtain the mean capillary velocity $U_c$, which is shown in Fig. \ref{fig:Uc}. It is observed that this outward radial velocity increases during the evaporation process, particularly accelerating in the later stages of droplet evaporation when $\frac{h}{R}$ $\ll 1 $\cite{saroj2021magnetophoretic,akdag2021interplay}.

The the time averaged capillary velocity, $U_c$ shows an approximately linear increase with substrate temperature, and linear fitting of the experimental data yields:

\begin{equation}
    U_c \sim 0.136~T_s - 3.026
    \label{eq:Uc}
\end{equation}

Equation~\eqref{eq:Uc} captures the trend observed in the experimental measurements of the internal velocity obtained using the $\mu$-PIV method.The capillary number, $C_a=\frac{\mu_{nf}U_c}{\sigma}$, is estimated to be on the order of $10^{-8}$. Since $C_a \ll 1$, surface tension dominates over viscous stresses, ensuring that the droplet maintains a nearly spherical-cap shape, which is consistent with our experimental observations.

\subsection{Transition of deposition morphology}

The deposition morphology changes significantly with an increase or decrease in substrate temperature relative to the non-heated condition. To quantify this effect, we introduce a non-dimensional parameter, $\Pi = J_a M_a$, which represents the relative importance of evaporation rate and thermocapillary flow within the evaporating droplet.

The non-dimensional parameter is defined as:
\begin{equation}
    \Pi = J_a M_a
\end{equation}

The temperature differences scale as $\Delta T_s = 0.000147\,T_s^{2.59}$ and $\Delta T_{avg} \sim 0.000143\,T_s^{2.66}$. Substituting these relations, $\Pi$ reduces to:
\begin{equation}
    \Pi = 1.54\,T_s^{5.25}, \qquad \Pi_{ref} = 1.54 \times 26^{5.25}
\end{equation}

To compare different substrate temperatures, $\Pi$ is normalized using the non-heated case as a reference:
\begin{equation}
    \Pi_{rel} = \frac{\Pi}{\Pi_{ref}} = \left(\frac{T_s}{26}\right)^{5.25}
\end{equation}

\begin{figure}[htb!]
    \centering
    \includegraphics[width=1\linewidth]{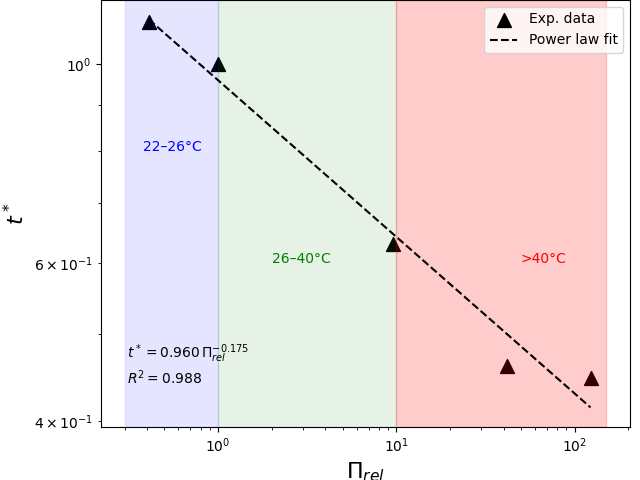}
    \caption{Variation of normalized peripheral width $t^*$ with $\Pi_{rel}$, showing a power-law decay and transition in deposition morphology.}
    \label{fig:Pi}
\end{figure}

The width of deposition pattern at the peripheral region is shown in Fig.~\ref{fig:thick}. A normalized width, $t^*$, is defined with respect to its value at $T_s = 26^\circ\mathrm{C}$. A plot of $\Pi_{rel}$ versus $t^*$ is used to characterize the transition in deposition morphology, as shown in Fig.~\ref{fig:Pi}. The power-law fit describing the variation of $t^*$ with $\Pi_{rel}$ is given by:
\begin{equation}
    t^* = 0.96\,\Pi_{rel}^{-0.175} \qquad (R^2 = 0.988)
\end{equation}

\subsubsection*{\textit{For $\Pi_{rel} \le 1$:}}
In this regime, coffee-ring formation is observed, as shown in Fig.~\ref{fig:deposition} for $T_s = 22$--$26^\circ$C, where $M_a$ is of the order $\mathcal{O}(10^3)$. Although Marangoni flow is present and induces thermocapillary circulation, its influence remains relatively weak compared to capillary flow. Particles near the contact line are transported toward the periphery and deposit there, while particles in the interior region move toward the center, as observed in Fig.~\ref{fig:piv}. However, during the late stage of evaporation, when $h/R \ll 1$, capillary flow dominates over thermocapillary flow, resulting in strong outward transport and the formation of a peripheral coffee-ring deposit.

With decreasing substrate temperature ($T_s = 26$ to $22^\circ$C), the droplet remains relatively warmer than the substrate, leading to a reduced evaporation rate (see Fig.~\ref{fig:J}) and consequently a thicker coffee ring. The slight particle deposition at the center is attributed to the near-zero velocity region at the droplet core (see Fig.~\ref{fig:piv}).

\subsubsection*{\textit{For $1 < \Pi_{rel} \le 10$:}}
In this regime, thermo-capillary flow strengthens with the increasing evaporation rate by increasing substrate temperature to $T_s = 40^\circ$C (see Fig.~\ref{fig:J} and Fig.~\ref{fig:Ma_Gr}). The $M_a$ is on the order of approximately $4.3 \times 10^3$. The coffee-ring formation persists. This behavior is associated with the condition during late stage evaporation, $h/R \ll 1$, under which capillary flow continues to dominate over thermo-capillary flow. The outward radial flow compensates for the enhanced evaporation near the contact line, transporting particles toward the edge and forming a relatively narrow peripheral ring. Readers are referred to the supplementary video provided with the manuscript.

\subsubsection*{\textit{For $\Pi_{rel} > 10$:}}
Further increasing the substrate temperature to $T_s = 53^\circ$C results in the formation of a dual-ring deposition pattern. This transition is associated with the strengthening of thermocapillary flow in the range $4.5 \times 10^3 \lesssim M_a \lesssim 7 \times 10^3$. The peripheral region of the droplet dries earlier than the apex region, as shown in Fig.~\ref{fig:camera} for $T_s = 53$ and $65^\circ$C, due to enhanced evaporation near the contact line. In this thin region, the thermal conduction resistance is minimal, facilitating efficient heat transfer from the substrate to the liquid. As a result, the interfacial temperature approaches the substrate temperature ($T_{int} \approx T_s$), leading to rapid evaporation and early particle deposition near the edge.

With further increase in substrate temperature to $T_s = 65^\circ$C, $M_a$ approaches $\mathcal{O}(10^4)$, resulting in a significant suppression of the coffee-ring effect.

Thus, $\Pi_{rel}$ serves as an effective scaling parameter to predict the transition from classical coffee-ring deposition to suppressed or multi-ring patterns governed by thermocapillary effects..
\subsection{Mechanism behind the peripheral deposition structure}
In contrast to the above observations, the deposition structure at the peripheral region differs significantly for $T_s \le 26\,^\circ\mathrm{C}$ and $T_s > 26\,^\circ\mathrm{C}$. For $T_s \le 26\,^\circ\mathrm{C}$, a framework of irregular polygonal tessellation network structures is observed, whereas for $T_s > 26\,^\circ\mathrm{C}$, a classical coffee-ring deposition accompanied by radial cracks is obtained.

The attractive capillary force between particles and the associated pressure difference arising due to the liquid meniscus during the late stage of drying can be expressed as
\begin{equation}
    F_{cap} = 2\pi \sigma r_p \cos\theta, \qquad 
    \Delta P = \frac{2\sigma \cos\theta}{r_m},
\end{equation}
where $F_{cap}$ is the capillary force acting on the particles, $\Delta P$ is the capillary pressure between particles, $\sigma$ is the surface tension, $r_p$ is the particle radius, $r_m$ is the radius of the meniscus connecting two particles, and $\theta$ is the contact angle.

For $T_s \le 26\,^\circ\mathrm{C}$, the evaporation rate is relatively slow, resulting in low particle velocities and sufficient time for particle rearrangement. During the late stage of drying, the progressive decrease in $r_m$ enhances both $F_{cap}$ and $\Delta P$, promoting gradual particle aggregation and redistribution. This facilitates the formation of an interconnected network of particle-rich channels, giving rise to irregular polygonal tessellation network structure.

In contrast, at higher substrate temperatures ($T_s > 26\,^\circ\mathrm{C}$), the evaporation rate increases significantly, leading to higher particle velocities and reduced time for structural rearrangement. Additionally, enhanced Marangoni convection further intensifies internal flow. As a result, rapid particle accumulation at the contact line and increased stress buildup suppress the formation of irregular polygonal tessellation networks, leading instead to a classical coffee-ring deposition accompanied by radial cracking.
\subsection{Onset of Marangoni instability and mechanism of central deposition pattern }
\begin{figure*}[htb!]
    \centering
    \includegraphics[width=1\linewidth]{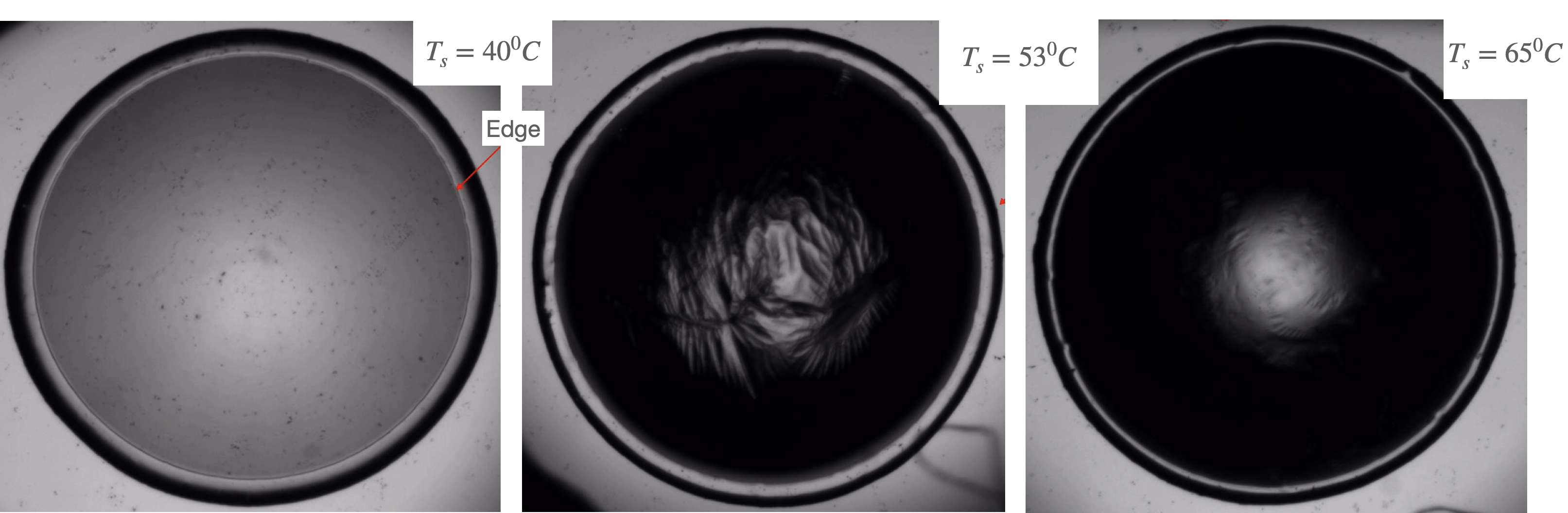}
    \caption{Top-view images illustrating interfacial deformation induced by Marangoni effect at higher substrate temperatures ($T_s$), highlighting the onset of Marangoni instability.}
    \label{fig:confo_shape}
\end{figure*}

Fig.~\ref{fig:confo_shape} presents the top-view visualization images captured using a confocal microscopy system. At $T_s = 53^\circ$C and $65^\circ$C, noticeable shape deformation of the droplet surface is observed. This shape deformation is likely caused by the Marangoni stress, which arises from surface tension gradients along the liquid--air interface. Corresponding side-view visualization images are shown in Fig.~\ref{fig:camera}. These images reveal that the peripheral region of the droplet dries earlier than the apex due to the higher evaporation rate near the contact line. Infrared (IR) imaging further confirms that the temperature near the edge region is higher, resulting in accelerated evaporation and consequently enhanced peripheral deposition.

In contrast, at $T_s = 40^\circ$C, no noticeable surface deformation or central deposition is observed. This indicates that the thermo-capillary flow at this temperature is not strong enough to overcome the capillary flow driven by the higher evaporation rate at the droplet periphery during late stage of evaporation. Therefore, $T_s = 40^\circ$C can be considered as the critical substrate temperature, above which dual-ring formation and suppression of the coffee-ring effect begin to occur. Additionally, particle agglomeration at higher substrate temperatures becomes evident, particularly in the central deposition region. For a clearer visualization of the internal flow dynamics and deposition evolution, readers are referred to the supplementary video provided with the manuscript. The morphology of the particles deposited in the central region mimics the structure of the deformed droplet interface. This suggests that the final deposition pattern retains the imprint of the interfacial deformation occurring during the evaporation process. Therefore, the particle assembly in the central region reflects the transient deformation of the liquid–air interface. The onset of Marangoni instability occurs during the later stage of  evaporation, as evidenced by both the side-view images (Fig.~\ref{fig:camera}) and the top-view observations (Fig.~\ref{fig:confo_shape}). Accordingly, $Ma = 4.5 \times 10^{3}$ is identified as the critical Marangoni number at which Marangoni convection transitions to Marangoni instability.
Overall, increasing the substrate temperature enhances both the evaporation rate and the Marangoni convection, while buoyancy-driven flow remains comparatively weak. The resulting deposition patterns are therefore governed by the competition between capillary flow and thermo-capillary flow, which determines the particle transport toward the edge and the final deposition morphology.

\section{Conclusion}
The present study systematically investigates the evaporative cooling and the resulting deposition patterns of a colloidal droplet containing $Al_2O_3$ nanoparticles on a hydrophobic glass substrate heated at different temperatures. The temporal evolution of the droplet height, contact angle, and volume is expressed as a function of the evaporation time using curve fitting and appropriate scaling arguments. The evaporation rate is found to increase with increasing substrate temperature. Infrared thermography is employed to measure the interfacial temperature distribution along the liquid–air interface. The temperature is observed to be higher near the contact line and decreases quadratically toward the apex region, resulting in the strongest evaporative cooling at the apex of the droplet. Based on these measurements, appropriate scaling arguments are developed to obtain a master interfacial temperature profile applicable for different substrate temperatures. Furthermore, an effective cooling efficiency is defined and evaluated, demonstrating enhanced cooling effectiveness at higher substrate temperatures. The evaporation flux is derived by assuming that heat transfer is primarily governed by conduction through the substrate, while the absorbed latent heat during evaporation leads to interfacial cooling. The final deposition patterns formed after complete evaporation exhibit the classical coffee-ring structure for non-heated or cooled substrates, with pronounced particle accumulation near the contact line. At lower substrate temperatures, a framework of interconnected particles forming a irregular polygonal tessellation network structures is also observed. As the substrate temperature increases up to $T_s = 40^\circ$C, a well-defined coffee-ring pattern with dense particle deposition at the edge is formed, accompanied by a reduction in the deposition width due to accelerated evaporation near the contact line. The development of Marangoni stresses along the interface leads to deformation of the droplet interface and induces internal circulation driven by thermo-capillary flow. At higher substrate temperatures, a dual-ring formation along with partial suppression of particle deposition at the edge is observed, indicating the increasing dominance of thermo-capillary flow. $\mu$-PIV measurements are performed to quantify the velocity distribution inside the droplet. The internal flow velocity increases with increasing substrate temperature, which is consistent with the predictions obtained from the scaling analysis.



\section*{conflict of interest}

There is no any conflict of interest



\nocite{*}
\section*{References}

\bibliography{aipsamp}

\end{document}